# A New Communication Theory on Complex Information and a Groundbreaking New Declarative Method to Update Object Databases

Heikki Virkkunen, Email: hvirkkun@gmail.com, Date: 5 April 2016

This article as a Blog:
http://hvirkkun.blogspot.fi/2016/04/a-new-communication-theory-on-complex.html

Download Java demo Implementation of the embed method as a Zip file:
https://drive.google.com/file/d/0B8eKvcE21R9ReGpKbWwzUXpLX2M/view?usp=sharing

In this article I introduce a new communication theory for complex information represented as a direct graph of nodes. In addition, I introduce an application for the theory, a new radical method, "embed", that can be used to update object databases declaratively.

The embed method revolutionizes updating of object databases. One embed method call can replace dozens of lines of complicated updating code in a traditional client program of an object database, which is a huge improvement. As a declarative method the embed method takes only one natural parameter, the root object of a modified object structure in the run-time memory, which makes it extremely easy to use.

From a Java program the embed method is called in the following way:

**db.embed(s);**

where s is the root object of a modified object structure in the run-time memory.

The communication theory behind the embed method states that modified complex information represented as a directed graph of nodes can always be transferred back to its original system in an exact and meaningful way. The theory can also have additional applications since today information often has a network or directed graph like structure and it often evolves or it is maintained.

The embed method applies the communication theory by modeling the object database and a modified object structure in the run-time memory as directed graphs of nodes without any artificial limitations. For example, different kinds of circular substructures and their modifications are allowed. Persistence in the object database is defined in a well known and accepted way, by reachability from persistent root objects. The embed method does not only transfer the modified information into the object database but also removes structures of garbage objects from the object database if any appear during the update operation, leaving the database in a consistent state. Garbage objects may appear if some objects in the object database lose references to themselves during the update operation. The embed method applies reference counting techniques. It understands local topology of the object database, avoiding examining unrelated objects in the database. For these reasons, the embed method is efficient and it scales for databases of different sizes.

At more abstract level the embed method can be understood as an updating method in the distributed system of two memories or directed graphs.

## 1. Introduction

It is well known that manipulating even complex object structures in a run-time memory is quite an easy task for a computer program; relationships of objects and contents of scalar fields of objects are easy to manipulate in the run-time memory. If a programming language offers automatic garbage collection, this manipulating task is even easier. But if object structures reside in a remote memory system, like in a persistent store, in an object database, manipulating or updating them is a much more difficult task, especially in complex cases.

When a client program updates an object database it typically first loads some object structure s, that has a root object s, from the object database. Then it modifies the loaded object structure in the run-time memory. After that, modifications in the object structure should be updated back to the object database. This updating task can be feasible if modifications of the object structure are not too complex. But if modifications are complex also the update operation is complex. In this article I call this kind of an updating task, in its widest meaning, a general update task/problem.

In an ideal case we would have a single method with simple interface which could solve this kind of an general updating problem, regardless of the complexity of the object data structures and their modifications. The single natural parameter for the method would be the root object of a modified object structure because it naturally defines the modified object structure in the run-time memory. The modified object structure could also be multi-rooted, but at the moment we can ignore it.

Current object database systems do not provide a command that could solve the general update problem. Instead, with them solving the problem may require several lines of complex updating code in a client program of the object database. That code can contain many interactions with the object database, not only update operations but also search and delete operations, and in addition complex programming logic. This is tedious and error-prone from a programmer's point of view and the same kind of effort must be repeated in different cases that resemble, more or less, each other. Indeed, difficulties in maintaining persistent object structures often form a serious bottleneck when designing and implementing useful application software. With SQL databases and their lower level table API difficulties are even more serious.

In my understanding, difficulties around the general update problem arise mainly from three sources.

First, modifications of an object structure in the run-time memory can be totally arbitrary and complex. For example, a client program can change the overall topology of the run-time object structure by removing objects from it or by adding new objects to it, some of them created by the client program, some of them loaded originally from the database. In addition, scalar fields of objects can be modified. For these reasons, a modified object structure in the run-time memory can be very different compared to what is currently in the object database. In the most common case, the modified object structure in the run-time memory can be any object structure, a network or graph of interrelated objects, consisting of arbitrary objects of which some are originally loaded from the database and some created by the client program.

The second difficulty in the general update problem arises from garbage objects in the update operation. Garbage objects may easily appear during an update operation as the short example below shows. If objects in the database can be shared between several users it makes the problem of determining garbage objects even more difficult. Naturally, a client program should not remove an object from the object database if someone else is referring to it. But this is often very difficult to determine. If potential garbage objects form complex structures, containing for example circular substructures, the problem is even harder. Therefore, it may be very difficult for a client program to determine if it can remove a useless-looking objects from the object database.

The following example shows that garbage objects may appear during a general update operation. Generally speaking, garbage objects may appear if some object p' in the object database loses a reference to itself during the update operation. Then p' and reachable objects from it become possible garbage objects. In this example we assume that an object database contains an object structure A → B → C (object A refers to object B which is refers to object C) of three objects which a client program loads in the run-time memory. After this the client program modifies the object structure in the run-time memory by constructing an object structure A → D → C. Here, object D is created on-fly by the client program. The client program may also have edited some scalar fields of objects in the structure. After this, the client program updates the object database with the modified object structure A → D → C. Obviously, after the update operation, the object database will contain the object structure A → D → C. But what happens to the object B in the database? It depends on whether someone else is referring to object B. If someone is referring to B, it will remain in the database. But if no one is referring to B, it is garbage and it should be removed from the object database during the update operation. This example was a simple one. Complex update cases can produce arbitrary complex structures of garbage objects.

The third difficulty in the general update problem is a meta level question; how an update method should update the database in a general case, i.e. is there always an exact and meaningful goal for an update operation? Perhaps inability to answer this question positively, or even formulate the question, is one reason why the general updating problem has not been considered widely in the literature or in the Internet. On the other hand, much discussion has been devoted to nearby concepts relating for the persistence of objects, for example in the contexts of object databases, NoSQL databases and object-relational mapping (O/R mapping). However, it turns out that an update operation always has a well defined goal and that it can even be achieved efficiently both algorithmically and in practice.

Of the current object database systems at least the object database system db4o [1] takes a step to the right direction in solving the general update problem. This database system offers a store command that takes the root object of the modified object structure as its single parameter and stores the modifications into the object database. However, that store command is not able to solve the problem fully. It cannot do the most difficult part, removing automatically garbage objects from the object database, if any appear during an update operation.

The general update problem can be approached by modeling both the object database and the modified object structure in the run-time memory as directed graphs of nodes. This is a natural way to model them and also a natural representation for complex information.

Let G0 denote the directed graph that represents the object database and G1 denote the directed graph that represents a modified object structure in the run-time memory. Now the first question in the general update problem is, how to define an unambiguous and meaningful result for an update operation where modified information in G1 is transferred back to its original system G0. This is perhaps a fundamental information or communication theoretical question for complex information represented as directed graph of nodes. In this paper I will answer this question with a communication theory that defines the result of the transferring operation.

A few things must be defined before it is possible to answer the general updating question. First, we must know which nodes in G1 are originally loaded from G0 and for such nodes we must know their corresponding nodes in G0. Here it is naturally assumed that two different node instances in G1 do not have the same corresponding node in G0. In addition, persistence in G0 has to be defined somehow to make possible to determine which nodes remain in G0 after an update operation, and which should be removed as garbage. Perhaps a null definition for the persistence in G0 is that all nodes in G0 are persistent. However, for our purposes persistence in G0 is defined in a well known and accepted way, by reachability; a node in G0 is persistent if it is reachable from one of the nodes marked as a persistent root nodes in G0. Otherwise a node in G0 is garbage and it should be removed automatically from the object database during the update operation. In object databases, root objects of the user's data structures are often natural candidates for the persistent root nodes (objects). For example, if a user has stored a tree like object structure “car”, with all its parts, in the object database, it is natural to assume that the car object itself is a persistent root object of the whole object structure representing the car.

The declarative embed method which I describe solves the general update problem. It achieves the results of the communication theory and it does it efficiently. The embed method takes the root object of a modified object structure as a single parameter and stores the modifications into the object database. In addition, if structures of garbage objects appear during the update operation, it removes them automatically from the database, leaving the database in a consistent state. The embed method and the communication theory behind it do not set any restrictions for complexity of the object database or complexity of the modified object structure in the run-time memory, or its modifications. For example, the different kinds of circular substructures and their modifications are allowed.

The embed method should be understood in wider contexts than only in the context of object databases. In the context of distributed memories the embed method translates complex modifications in the remote memory (for example the object database) into simple modifications in the local memory (for example the run-time memory). In its most abstract level, the embed method can be understood in the context of the distributed system of directed graphs.

The embed method consists of two co-operating phases, an update phase and a garbage collection (gc) phase which are executed subsequently. The embed method stores reference count information in the objects in the object database and keeps that information up to date. After the embed method has executed the update phase, the content of the modified object structure exists in the object database. The update phase also collects information about possible garbage objects in the object database. The garbage collection phase utilizes that information and applies reference counting techniques. The garbage collection phase understands local topology of the object database, avoiding examining unrelated objects in the database. As a result, the embed method is efficient and scales for databases of different sizes. The embed method may also apply to distributed databases, for example in the Internet, because the reference mechanism between the nodes is not relevant in the algorithm.

The embed method can be tested with the published Java demo implementation:

https://drive.google.com/file/d/0B8eKvcE21R9ReGpKbWwzUXpLX2M/view?usp=sharing

Users can use this implementation for arbitrary complex update tests. Due to the clear background of the embed method, the size of the implementation is only about a thousand lines of Java code. That includes also utility methods. The implementation uses the SQLite database as an underlying lower level raw data store. Therefore, the implementation is formally an object-relational database. However, the underlying data store could be, for example, a pure random access file. Together with the demo implementation comes some example applications that demonstrate the power and generality of the embed method.

**The organization of the article:**

## *2. Structure of the Object Database*

The embed method treats the object database as a directed graph of objects or nodes. There are no restrictions for the structure of this graph. For example, the graph may be disconnected or connected and different kinds of circular substructures are allowed. Real object databases are probably more often disconnected than connected.

Each non-null object in the object database has a unique **id**, an integer greater than zero. Objects in the database refer to other objects in the database with these **id** values in their pointer fields. An object can also refer to itself. Null objects are not stored explicitly in the database; value zero in a pointer field represents reference to a null object , i.e. a null reference. It is assumed that id values are not reused in the object database. In some systems, like in distributed databases, the **id** value can be more complex structure than a plain integer.

The object database defines when an object is persistent in the database, or when it is useless, i.e garbage. In short, persistence in the database is defined by reachability from persistent root objects. This is a well known and accepted way to define persistence. Persistent root objects are persistent by themselves, independently of the references from the other objects in the database. It can be thought that someone is referring to a persistent root object from outside the database which makes it persistent. There can even be many such outer references to a persistent object. After this, persistence is defined by saying that an object in the database is persistent if it is reachable from some persistent root object. Other objects in the database are garbage and they should be removed from the database an update operation. Note that a persistent root object is also persistent because it is reachable from itself.

Objects in the database keep track of the number of references to themselves in two reference count fields. The first reference count field is the outer reference count field, the **orc** field, and the second reference count field is the internal reference count field, the **irc** field. The **orc** field defines how many references exist to an object from outside the object database. If **orc** of an object is greater than zero then that object is a persistent root node. Some systems may allow only **orc** values one and zero. The **irc** field defines how many references exist for an object from other objects in the database, including the object itself. Obviously, an object is garbage if **irc**=**orc**=0 for it. However, an object may be garbage even if its **irc** is greater than zero. This happens if garbage objects form circular structures.

When a client program loads a non-null object (lonely or as a part of larger structure) from the object database the id value of the object is loaded at the same time and it is bound to the loaded object in the run-time memory. On the other hand, when the embed method allocates for a non-null run-time object having id=0 (i.e. for an object that does not yet have a corresponding object in the database) a corresponding object into the object database, the run-time object gets the id of the allocated object. When a client program creates a run-time object from scratch it automatically gets the id value zero in the run-time memory. The id value zero defines that an object does not have yet a corresponding object in the database.

The **id** value can be bound to a run-time object in various ways. In the demo implementation of the embed method it is simply assumed that a run-time object has an extra **id** field of the type Java int for this purpose. In real systems it may be useful to implement the **id** field outside the object. In Java environments this can be done, for example, by using weak reference techniques. With weak reference techniques it is possible to bind any information to a run-time object without polluting the structure of the object or affecting the lifetime of the object.

## 3. Two Examples Demonstrating Results of the Embed Method

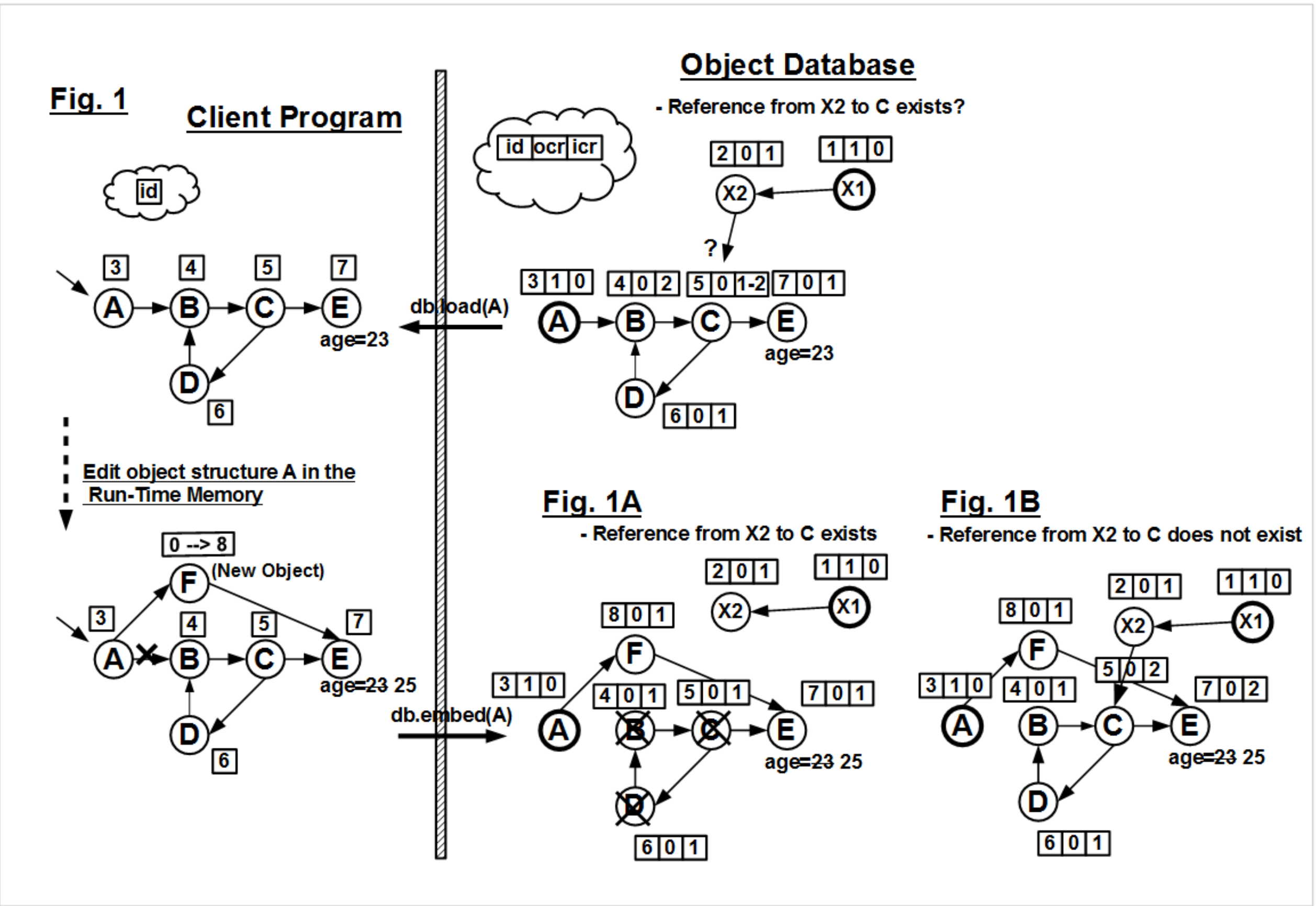

**Fig.1** shows two related update examples of how the embed method works.

The purpose of these two examples is to describe that regardless of the modifications of the run-time object structure and the content of the object database the update method embed always has an unambiguous and meaningful result. These examples also show that local topology of the object database, the related graph, affects the result of the update operation.

In the abstract level these examples describe that modified complex information represented as a directed graph of nodes can always be transferred back to its original system in an exact and meaningful way.

To demonstrate the real embed method, **Fig. 1** shows contents of **id**, **orc** (outer reference count) and **irc** (internal reference count) fields of the objects in the database objects and contents of the **id** fields bound to run-time objects.

In **Fig. 1** the client program first loads object structure A from the object database with a command ”db.load(A)”, the command provided by the object database system. The load command loads object A and all reachable objects from it. Note that the object database contains two persistent root objects, object A and object X1, which both have **orc**=1. In the object database there is an optional reference from object X2 to object C. However, the existence of this reference does not affect the result of the loading operation. Note that loaded object structure A is "complex". For example, it contains a circular substructure, B → C → D → B. Note also that the **id**s of the objects are loaded within the objects.

After loading object structure A the client program modifies it in the run-time memory. First, the pointer field of A referring to object B is changed to refer to object F in the run-time memory. The client program has created object F from scratch. Therefore the **id** of F is zero. Then object F is set to refer to object E. In addition, scalar field **age** of E is updated to contain value 25.

Now modifications of object structure A in the run-time memory are ready. Note that the modifications changed the topology of object structure A. For example, circular substructure B → C → D → B disappears from structure A and node F was added to it.

Next, the client program updates the object database with modified object structure A. The client program makes the call:

```
db.embed(A);
```

What should happen?

Run-time structure A contains now only three objects, structure A → F → E. Therefore, the object database is updated only with this object structure. The garbage collection system, if that exists, of a programming language removes objects B, C and D from the run-time memory.

Clearly, object structure A → F → E should now exist in the object database. Therefore, the embed method creates object F into the object database and sets F to refer to object E in the database. In addition, scalar field **age** of object C gets the new value of 25 in the database. Furthermore, object A now refers to object F not to B any more. But what happens for object B which loses a reference from object A, and for all objects reachable from object B, i.e. for objects B, C, D and E in the object database?

Clearly object E cannot disappear from the database, because it is still reachable from persistent root node A, through object F. But objects B,C and D are not any more reachable from persistent node A. We can notice that objects B,C and D are reachable from the other persistent root object, object X1, only if the reference from X2 to C exists in the object database.

**Fig. 1A** describes the result of the embed method when the reference from X2 to C does not exist. In this case objects in circular object structure B → C → D → B disappear from the database as garbage. Note that this happens even though all the nodes in this structure have internal reference count, **irc**, as one, i.e. greater than zero.

**Fig. 1B** describes the opposite situation, when the reference from X2 to C exists. In this case objects in circular object structure B → C → D → B do not disappear from the object database.

These two examples show that local topology of the object database, the corresponding graph, can affect heavily the result of the embed method. Note that in these examples it may be that the client program does not know anything about object X1 in the object database, not even its type.

During the update operation, the embed method must also update the reference counts of the objects in the object database accordingly as shown in **Figures 1A** and **1B**. In addition, when the embed method allocates object F in the object database it also binds its **id** to object F in the run-time memory.

## 4. How Modified Complex Information Represented as a Directed Graph of Nodes Is Transferred Back to Its Original System in an Exact and Meaningful Way

Let G0 be an original directed graph of nodes and let G1 be a modified copy of a subgraph of G0. For example, G0 can represent the object database and G1 can represent a modified object structure in the run-time memory.

Structures of the graphs G0 and G1 are not restricted in any way. For example, they can be multi-rooted, disconnected, or they can contain circular subgraphs. Because G1 is a modified copy of a subgraph of G0, some of its nodes (zero or more) are modified copies (containing zero or more modifications) of nodes in G0. If a node p in G1 is a modified copy of a node p' in G0 it is said that p' is a corresponding node for the node p. In most general case, G1 can be any graph consisting of nodes of which some are originally copied or loaded from G0 and some not. It is naturally assumed that the two nodes in G1 do not have the same corresponding node in G0. We denote nodes in G1 with letters p,q,.. and their corresponding nodes in G0 as p′,q′,.. , if a corresponding node exists.

Now, modified information in G1 is transferred back to G0. This is done in two phases which are the update phase and the garbage collection phase.

The update phase consists of four steps. The first three steps are:

**Step 1:**
Separate non-null nodes in G1 in two groups, white nodes and gray nodes. A node p in G1 is a white node if it does not yet have a corresponding node p' in G0. Otherwise, when p has a corresponding node p' in G0, p is a gray node. This step can be performed by traversing the graph G1.

**Step 2:**
For each white node p in G1 allocate the corresponding empty node p' in G0. Note that after this step also white nodes in G1 has their corresponding nodes in G0.

**Step 3:**
Update the content of G0 with the modifications in G1. Updating is done node by node for the non-null nodes p in G1. In updating the content of the node p replaces the content of p' in G0 in a flat way. It is not here very interesting algorithmically whether the nodes have a field structure and whether some fields are scalar fields. In pure mathematical graphs the nodes do not have a field structure and they do not contain scalars. Also, null references, references to null nodes, are not allowed in a pure mathematical graph. Nodes (objects) in the object database and in a modified object structure in the run-time memory have a field structure and fields can contain both reference and scalar values. In addition, reference fields can contain null references. When p' is replaced with the content of p, it can be thought that the content of p' is first cleared and then the content of p is copied to p' in a flat way. In copying, if p refers to a non-null node q in G1 reference to q' is copied to p'. Possible null references are copied as a null reference. Possible scalar values are copied as such.

The update phase, with its first three steps, is actually a simple operation. In some sense these steps lonely without the fourth step (described below) transfers formally modified information in G1 back into G0. It is a good question whether result of these three steps is known in a theoretical sense before and whether the significance of the result has been understood well in the context of updating object databases. The db4o database [1] provides the store operation which can be interpreted from this perspective.

Note that the update phase with its first three does steps not require any extra properties from the graphs G0 and G1. The only natural requirement is that there is a mapping from the node p in G1 to its corresponding node p' in G0, when the corresponding node p' exists. This does not require, for example, that the nodes should have some **id**s to make possible this mapping. However, the implemented embed method uses **id**s of nodes to form the mapping. In the implemented embed method the **step 2** includes also the binding of the **id** of the allocated node p' for the node p.

In the context of the implemented embed method it is interesting how the counts of incoming references, internal reference counts, of the nodes in G0 change as a consequence of the first three steps of the update phase because the implemented embed method keeps up to date those reference counts. However, these changes can be expressed without storing the reference count information or their changes in nodes or anywhere. In the implemented embed method reference count information is stored in the nodes in the database. Changes of internal reference counts can be defined in the situation when **step 2** of the update phase has been finished but before **step 3** has not yet been started:

-If p is a gray node in G1 and p' refers to a non-null node q' then the internal reference count of q' decrements by one. If p' refers to q' many times, then the internal reference count of q' decrements by one as many times.

-If p is a node (a white or gray node) in G1 and p refers to a non-null node q, then the internal reference count of the node q' increments by one. If p refers to q many times, then the internal reference count of q' increments by one as many times.

Phase 2 is a garbage collection. In this phase the garbage nodes that appear in the G0 as a consequence of the update phase, are detected and removed from G0. After the garbage collection phase has been performed, G0 is again in a stable state. It is naturally assumed that G0 was in a stable state before the update phase was performed.

Performing garbage collection requires that the persistence is defined somehow in G0. If G0 is an object database, persistence is defined by reachability from persistent root objects as described before. It is assumed that persistence in G0 is defined in this way.

Before performing the garbage collection, the fourth step of update phase is executed:

**Step 4:**
For selected white nodes in G1 mark their corresponding nodes in G0 as persistent root nodes.

This step can be explained with an example.

Suppose that you are storing, for example, a tree like object structure “car”, with all its parts, into the object database. Suppose also that the root object car itself is a new one, i.e. it is a white node. The car may consist of both old and new parts. It is now obvious that you want to make the object car', corresponding the root object car, in the object database a persistent root object to be sure that the whole car structure will be persistent in the object database. Hence, only the root node (object) car is selected in **step 4**.

On the other hand, if the car would be an old one, i.e the root object car would be a gray node, perhaps, no nodes would be selected in **step 4** because in this case the car' would be a persistent root object beforehand.

We can also imagine situations in which several white nodes are selected in **step 4**. This can happen, for example, if G1 is multi-rooted with several white root nodes. Then perhaps all white root nodes of G1 might be selected in **step 4**.

After **step 4** garbage collection is performed. It could be performed simply by traversing the whole G0 starting from its persistent root nodes. The nodes that are not reached from any persistent root node in G0 would be garbage and should be removed from G0. However, in practice, this is not an efficient algorithm for garbage collection. Therefore, the implemented embed method applies more efficient methods using, for example, reference counting techniques.

Together, the update phase and the garbage collection phase define a new communication theory that defines how modified complex information represented as a directed graph of nodes can always be transferred back to its original system in an exact and meaningful way.

Note: There could also be systems where persistence in G0 is defined in other way than described. For example, some kinds of circular structures and perhaps some reachable parts from them could be persistent in G0. Persistence could also be some kind of statistical phenomenon in G0 Perhaps environment of G0 could somehow define the persistence in G0. The garbage collection would, of course, follow the given rules of persistence. I will not to discuss this further here.

## 5. About the Algorithm and the Demo Implementation of the Embed Method

The algorithm and the demo implementation of the embed method are included at the end of this article. The algorithm is directly from the the demo implementation and both are written in the Java language. To make tests with the embed method, you must download the zip file containing the Java demo implementation:

https://drive.google.com/file/d/0B8eKvcE21R9ReGpKbWwzUXpLX2M/view?usp=sharing

The demo implementation can be used for arbitrary complex update tests. The zip file also contains some example applications that demonstrate the power and generality of the embed method.

The size of the demo implementation is only about a thousand lines of Java Code. If something is unclear it is quite easy to read the whole source code.

The implementation uses SQLite database as an underlying data store. Therefore, the implementation formally represents an object-relational (O/R) database, but that is not very important here. The underlying data store could be, for example, a pure random access file.

The demo implementation does not try to be efficient as an implementation. It is far from it. For example, fields of the nodes in SQL tables are updated one by one with SQL, and unchanged fields are also updated. In addition the utility structures are simple and inefficient for real purposes and too little intermediate data, like reference count information of the related nodes, is kept in the run-time memory. Of course, more advanced features, like, for example, cache or session have not been implemented. The implementation contains a simple search method which you can use to load object structures from the object database into the run-time memory.

The implementation supports two kinds of persistent nodes (objects), ordinary nodes (fixed nodes), and list nodes. These types of nodes are sufficient for many purposes. However, exact node types are not important from the algorithm's point of view. For simplicity, inheritance is not supported for the Java types of the nodes.

Fields of the nodes are either of type scalar or pointer. For simplicity, the allowed scalar types are only the Java types int, Integer and String. In the lists the scalar type int is replaced with the scalar type Integer because, in the run-time memory Java, ArrayList is used as an underlying structure.

In the run-time memory, the type of the field of the fixed node is determined from the field but the type of the field of the list node is determined from the content of the field. The list node is very flexible because the same list node can contain items of different types and, of course, the size of the list can vary. For example, one list node can contain scalar values and pointer values, and pointer values can refer to fixed nodes or to list nodes. To avoid ambiguity, in the run-time memory a null value in a list node is interpreted as a special value “None”. In the database fields of the list nodes have explicit types.

One useful node type for practical purposes could be a map, where the key would represent the name of the field and the value would represent the value in the field. However, this node type has not been implemented in the demo implementation. On the other hand, fixed nodes could be implemented as maps in the underlying data store, because field structure of a fixed node can vary in different releases of a client program and the database. In this scenario, when a node is read from the database, only the existing fields of a fixed node in the run-time memory are filled, and if the field of the run-time node does not exist in the database that field is not filled in the run-time node, or it is filled with some default value like null.

Below are two examples (Friend and Book) of Java types of fixed nodes. Also the Java type of the list node is presented.

Naturally the client program defines the Java types of the fixed nodes it uses. The Java type of the list node, ListNode, is defined by the system because it is the same for all client programs. In these examples, fields of the nodes Friend and Book are public, but this is not a requirement. Every node (object) in the run-time memory must contain the **id** field which identifies the corresponding node in the object database, when such a node exists. If the corresponding node does not exist, the **id** field contains the value zero. As mentioned before, in real systems the **id** field could be implemented outside the object by using Java weak reference techniques.

```
public class Friend
{
  public int id; //Compulsory id field.
  public String name;
  public Integer age;
  public Book book; //Reference to a fixed node of type Book.
  public ListNode list;  //Reference to a list node.
  public Friend(){} //Default constructor required.
  //Other methods..
}


public class Book
{
  public int id;
  public String name;
```

```
  public int price;
  public Book(){}
  //Other methods..
}

//The database defines the run-time type of the list node.
//The Java ArrayList is used as an underlying structure.
public class ListNode
{
  public int id;
  public ArrayList<Object> list = new ArrayList<Object>();
  public ListNode(){}
  //Other methods..
}
```

## ***6. Notes***

Before reading this section, it is good to read and understand the algorithm of the embed method.

- Let s be a modified structure in the run-time memory. Suppose that all the modifications in s are in its substructure p. For example, s may be a large tree and p a smaller subtree of s. In this case it is more efficient to call the embed method for p than s to avoid to examine the whole s in the embed method. Generally, the embed method can be called for an any node (object) in a modified object structure s.

- In real systems, it may be a good idea that the database system performs the garbage collection phase of the embed method to avoid extra traffic between the client program (a library) and the database.

- The update phase of the embed method can be implemented recursively in a such way that nodes in s, in run-time memory, are traversed only once.

- When the client program calls the embed method, it can know strategic nodes in the database (their **id**s), which are not garbage. Therefore, the embed method could take as an extra parameter a list of **id**s (or run-time nodes having those **id**s) of nodes which are not garbage in the database. The garbage collection phase of the embed method could then use these **id**s to stop walking in the graph Z (see the algorithm) when determining garbage nodes. This could make the embed method even more efficient in the appropriate situations. Of course, the client program should be very careful when defining the list of **id**s non-garbage nodes. This technique could be used, for example, in the embedded systems but also for other purposes .

- It is possible to add extra outer reference count fields for the nodes in the database. For example, one such field could be for an administrator who could define that some nodes in the database are persistent independently of the decisions of the users of the database. Perhaps in this case the extra reference count field could contain only value zero or one. Another type of system field could prevent updating a node if a user is not authorized for that operation.

- A “raw delete” method invoked like “db.forceDelete(p)” could be a useful and powerful. It could be invoked for any node p that has a corresponding node p' in the database. This method would remove the node p' from the database regardless of whether someone is referring to it. In addition it would remove garbage nodes caused by the removing of the p' from the database. This method could be implemented in the following way. First, lower the reference counts of non-null child nodes of the p' accordingly and put their ids in the set seedGarbageIds. After that, remove the node node p' from the database. Then call the garbageCollection method. The database can now contain dangling pointers which are referring to p' (and perhaps other dangling pointers). However, dangling pointers are easy to identify in the database (In the run-time memory the situation may be opposite) Therefore dangling pointers can be interpreted as null references. On the other hand, dangling pointers could be even replaced with null references on the fly when they are seen by the system. Generally, ability to identify dangling pointers in the object database can be an important architectural property.

- It is possible to imagine a system with only one reference count field where the reference count field contains the sum of the outer reference count and the internal reference count. Ability to interpret dangling pointers as null references may be a helpful feature in this implementation.

- It is easy to implement a multi-root version of the embed method. This does not affect the algorithm essentially. The Java invocation of this kind of embed method could be “db.embed(s1,s2,..,sn)” where a modified object structure consists of all the nodes reachable from the nodes s1,s2,..,sn. The modified object structure could be connected or not.

- The embed method in a demo implementation, its update phase, applies the following handlings for the nodes p in a modified object structure s in the run-time memory:

  -If p has **id** value zero (p is a white node) save p, otherwise, when **id** of p is greater than zero (p is a gray node) update the corresponding node p' in the database with p.

  -If p is the root node s and **id** of p is zero (p is a white node) make the corresponding node p' a persistent root node in the database.

  It is easy to define other useful handlings for the nodes p in s. For that, the implementation of the update phase of the embed method should be modified accordingly. This task is not impossible for a qualified programmer, after understanding the basic embed method in the network level. A handling can be bound for a node p in the same way as the **id** field of the node, for example, by using weak references techniques in the Java systems. A handling can have a default post state in the embed method, i.e. how the node p will be handled in a next call of the embed method. Some handlings fit for the fixed nodes, some for the list nodes and some for the both.

  Here are some example handlings, or families of handlings, which can be useful:

  -Suppose that **id** of p is greater than zero. If p' exists in the database, update p' with p. Otherwise save the node p. In this handling it is not an error that a node that has **id**>0 does not have a corresponding node in the database.

-Pass a node p. This handling applies perhaps best for nodes having **id**>0, i.e. for nodes having their corresponding nodes in the database. In this case the update phase does not update p' but continues from p to its descendants. This handling is useful, for example, if s consists of many nodes that have **id**>0 but only few of them, which are dispersed in s, have to be updated. It is also possible to define "passing" in the way that, for example, only defined set of scalar fields in p are updated.

-Update the defined set of fields of the node defined by p but do not continue through p (when collecting nodes of s in update phase) to its descendants. The set of fields can be an any defined set of fields. The node p can define the node in the database in alternative ways, with the **id** of p or with other content of p. This family of handlings can be useful, for example, if some subtrees in s need to be updated. If p defines the node in the database with another way than with the **id**, then it may be useful that afterward p gets **id** value of the defined node in the database.

- It is possible to implement a transaction version of the embed method. First beginTransaction is called and then embed method (and perhaps some other database methods) is called several times and then finally endTransaction or rollBack is called. For example:
  ```
  beginTransaction();
  ..
  embed(s1);
  ..
  embed(s2);
  ..
  embed(s3);
  ..
  endTransaction();
  ```

  Inside a transaction the embed method must behave a little differently than in the demo implementation. In the transaction case the embed method calls only the update phase (with some quite small modifications), not the garbage collection phase. Then the method endTransaction calls the garbage collection phase. In this way intermediate structures are not removed accidentally from the database during the transaction. Of course this requires a little programming work but it should not be impossible for a qualified programmer.

- It is possible to solve a well known "general synchronization problem" with the ideas behind the embed method. Suppose that we have a master object database A in a server. Then we take a part of it into the smaller database B, for example into a laptop computer. After that the laptop computer is disconnected from the master database A. Then we make updates into the smaller database B in the laptop computer. Afterward we want to synchronize content of B back to the master database A. This can be done for example in the following way. In this implementation the nodes in B have two id fields, **idA** representing the id in the A and another id, **idB**, representing the id in B. When nodes, or node structures, are first loaded from A to B, all nodes in B get and **idA** into their both id fields. When the database B is updated and new nodes are added to B, they get **idB**>0 (in a normal way), and **idA**=0 to mark that they will be new to the master database A. Finally B is synchronized (embedded) back to the master database A. In this operation the whole content of B is handled like a modified object structure in the run-time memory, in the normal embed operation. The id field **idA** is used as id field in the embed operation. This example makes the embed method a transitive operation in the described way.

- Garbage collection phase of the embed method is easier to implement if we know that garbage nodes does not form circular structures in the database. It is likely that this is true in many traditional database systems. In these cases garbage collection can be implemented in the following simplified way. First, it is examined if the set seedGarbageIds contains a node p' (its **id**) having both reference counts, **icr** and **orc**, zero. If this is false no garbage nodes exist. Otherwise **id**s of non-null child nodes of p' are added to the set seedGarbageIds unless they are already there. At the the same time, internal reference counts of these non-null child nodes are decreased accordingly. Then the **id** of p' is removed from the set seedGarbageIds, and the node p' is removed from the database. The procedure is repeated until the set seedGarbageIds does not contain any node p' having **orc**=**irc**=0.

## 8. The Algorithm of the Embed Method

### 8.1 Global Data Structures

```
//The embed method is called for a modified object structure s
//(with its root node) in the run-time memory. In the beginning
//of the embed method non-null nodes in s are separated in two
//groups, white nodes and gray nodes. A non-null node p in s is
//a white node if its id is zero, i.e. if p does not yet have a
//corresponding node in the the database. Otherwise, if id>0, p
//already has a corresponding node in the database with that id
//and p is a gray node. White nodes are collected in the list
//whiteNodes and gray nodes are collected in the list GrayNodes.
ArrayList<Object> whiteNodes;
ArrayList<Object> grayNodes;

//As a side effect, the update phase of the embed method
//collects ids of potential garbage nodes, it finds, in the
//set seedGarbageIds. When the update phase has been finshed,
//all possible garbage nodes belong to set or graph Z
//consisting of the nodes reachable from the nodes having their
//ids in seedGarbageIds. It is possible that only some, or none
//of the nodes in Z are garbage. Z may also be empty.
HashSet<Integer> seedGarbageIds;

//The map countOfInternalReferencesInZ is used to count the
//internal references inside Z. The key is id of a node and the
//value is count of internal references to that node in Z.
HashMap<Integer,Integer> countOfInternalReferencesInZ;

//In Z, nodes which are referred to from outside the Z,
//cannot be garbage. Ids of those nodes are collected in the
//set idsOfNodesRefOutsideZ.
HashSet<Integer> idsOfNodesRefOutsideZ;
```

### 8.2 The Embed Method

```
//This is the embed method which a client program calls to
//update the object database with a modified object structure s
```

```
  //in the run-time memory. The method is called with the root
  //object s of the modified object structure. The embed method
  //consists of two co-operating phases, the update phase and the
  //garbage collection phase.
  public void embed(Object s)
  throws Exception
  {
    update(s);
    garbageCollection();
  }
```

## 8.3 Private Methods

```
  //The embed method calls the update method with the root node s
  //of a modified object data structure in the run-time memory.
  //After the update method has been fineshed, content of the
  //object structure s exists in the database.
  //
  //As a side effect, the update method produces the set of ids of
  //potential garbage nodes, it finds, into the set seedGarbageIds.
  //The set seedGarbageIds is input for the garbage collection
  //phase of the embed method.
  //
  //Below the terms "step 1", "step 2", "step 3" and  "step 4"
  //refer to the steps described in the article, in section 4.
  private void update(Object s)
  throws Exception
  {
    //Global lists for the white nodes and the gray nodes in s.
    whiteNodes = new ArrayList<Object>();
    grayNodes = new ArrayList<Object>();

    //This boolean value dscribes whether the root node s is
    //white. This information is used in the step 4 of this
    //method.
    boolean rootIsWhite =
      getId(s) == ID_ZERO;

    //Create the global set seedGarbageIds.
    seedGarbageIds = new HashSet<Integer>();

//Step 1 and step 2 of the update method:
    //Step1: Collect white nodes in object the structure s in the
    //list whiteNodes and gray nodes in s in the list grayNodes.
    //Step 2: For each white node in s: Allocate the same type of
    //empty node in the database.
    collectWhiteAndGrayNodes(s);

//Step 3 of the update method:
    //Handle changes of internal reference counts of the nodes
    //caused by updating the database with the white nodes.
    handleReferencesFromWhiteNodesInDB();

    //Handle changes of internal reference counts of the nodes
    //caused by updating the database with the gray nodes.
    handleReferencesFromGrayNodesInDB();

    //Update the database with the white nodes.
    copyContentsOfWhiteNodesToDB();

    //Update the database with the gray nodes.
    copyContentsOfGrayNodesToDB();


//Step 4 of the update method:
    //For selected white nodes in s make their corresponding nodes
    //in the database persistent root nodes by setting them orc=1.
    //
    //We simply select only the root node s, if it is a white node.
```

```
  //Otherwise no nodes is selected.
  //
  //When a root node s is a white node it is obvious that the
  //user wants to make s' a persistent root node to make s' and
  //all the rechable nodes from it persistent in the database.
  //
  //There exist rare cases where a user may want that for a
  //white root node s the corresponding node s' gets orc=0 and
  //does not not become a persistent root node in the database.
  //For example if the run-time object structure s is a circular
  //structure s -> p -> s, where s is a white node and p is a
  //gray node, a user may want that persistence of s' depends on
  //the persistence of p'. Therefore orc could be zero for s in
  //this case.
  //
  //If the embed method is called for a gray node s, then the
  //orc of s' is kept as it was. This is a natural solution.
  //However, this can cause a strange but perhaps correct
  //effects in some cases; the node s', and perhaps some
  //reachable nodes from it can disappear from the object
  //database as aconsequence of an executed embed method. The
  //following example describes this.
  //
  //Let s' originally refer to some node p' and p' refer back to
  //s' and suppose that no-one else in the database is referring
  //to s', and that orc=0 for s'. After that circular structure
  //s -> p -> s is modified in the run-time memory by setting p
  //to refer to a null node, i.e. not to s anymore. After that
  //the embed method is called for s. As a consequence s'
  //becomes garbage, because now irc=orc=0 for s'.
  if (rootIsWhite)
    incrORC(getId(s));

  //Free the lists reserved for white nodes and gray nodes.
  whiteNodes = null;
  grayNodes = null;
}


//Method collectWhiteAndGrayNodes is called from the method
//update. This method separates white nodes and gray nodes in the
//object structure s. White nodes are collected in the list
//whiteNodes and gray nodes in the list grayNodes.
//
//In addition, the method allocates for each white node p in s a
//corresponding empty node p' (having a corresponding type) in
//the object database. The node p gets the id of p' into its
//id field.
private void collectWhiteAndGrayNodes(Object p)
throws Exception
{
  //Null nodes are note collected.
  if (p==null)
    return;

  //The same node instance is not collected twice.
  if (bufferContainsNode(whiteNodes,p)
      || bufferContainsNode(grayNodes,p))
    return;

  int id = getId(p);

  //If p is a white node, collect p and allocate a corresponding empty
  //node p' in the database. Assign id of p' to p.
  if (id == ID_ZERO)
  {
    id = allocateNodeInDB(p);
    setId(p,id);
    whiteNodes.add(p);
  }
  else //Collect a gray node p.
```

```
    grayNodes.add(p);

  //Collect non-null child nodes of p if not yet collected.
  ArrayList<FieldT> pointerFields = getFields(p.getClass(),p,1);
  for (FieldT f : pointerFields)
    collectWhiteAndGrayNodes(f.value);
}


//Handle changes of internal reference counts caused by updating
//the object database with white nodes.
private void handleReferencesFromWhiteNodesInDB()
throws Exception
{
  for(Object p:whiteNodes)
    handleReferencesFromWhiteNodeInDB(p);
}
//Here p is a white node. Updating p' with the p can increase
//internal reference counts of some nodes in the database. These
//changes are updated in this method.
//
//The following rule gives the result:
//If a white node p refers with a pointer field f to a non-null
//node q then the internal reference count of the node q'
//increments by one. If p refers to q many times (with many
//pointer fields), then the internal reference count of q'
//increments as many times.
private void handleReferencesFromWhiteNodeInDB(Object p)
throws Exception
{
  ArrayList<Integer> C1 = getIdsOfNonNullChildNodes(p);
  for(Integer id:C1)
    incrIRC(id);
}


//Handle changes of internal reference counts caused by updating
//the object database with gray nodes.
private void handleReferencesFromGrayNodesInDB()
throws Exception
{
  for(Object p:grayNodes)
    handleReferencesFromGrayNodeInDB(p);
}
//Here p is a gray node. Updating p' with the p can change
//internal reference counts of some nodes in the database. These
//changes are updated in this method.
//
//The following rules give the result:
//
//1)
//If, before updating the node p' with a gray node p, the node
//p' refers with a pointer field f' to a non-null node q' then
//the internal reference count of q' decrements by one. If p'
//refers to q' many times (with many pointer fields), then the
//internal reference count of q' decrements as many times.
//
//2)
//If a gray node p refers with a pointer field f to a non-null
//node q then the internal reference count of the node q'
//increments by one. If p refers to q many times (with many
//pointer fields), then the internal reference count of q'
//increments as many times.
//
//The algorithm does incrementing/decrementing in such a way
//that the internal reference count of a node is only
//incremented or decremented, not both.
//
//If, after updating, the node p' does not any more refer to a
//non-null node q', then q' may be garbage. In this case the id
//of q' is added conditionally to set seedGarbageIds, if it is
```

```
//not yet there.
private void handleReferencesFromGrayNodeInDB(Object p)
throws Exception
{

  //Construct the list C1 = (id(q1),..,id(qn)) of the ids of
  //non-null child nodes of the node p. If p refers several
  //times to the same non-null child node q then the id of q is
  //as many times in the list C1.
  ArrayList<Integer> C1 =
    getIdsOfNonNullChildNodes(p);


  //Construct the list C2 = (id(q'1),..,id(q'm)) of the ids of
  //non-null child nodes of the node p' (before p' has been
  //updated with p). If p' refers several times to the same
  //non-null child node q' then the id of q' is as many times
  //in the list C2.
  ArrayList<Integer> C2 =
    getIdsOfNonNullChildNodesOfDBNode(getId(p));

  //The set I will contain the intersection of C1 and C2. (The
  //same id is not twice in I).
  HashSet<Integer> I = new HashSet<Integer>();

  //Make lists C1 and C2 disjoint. The intersection of C1 and C2
  //is collected in the set I. Note that C1 can contain the
  //same id several times, before and after making C1 and C2
  //disjoint. The same is true for the list C2.
  //For example: Let
  //C1 = (1,2,1,3,4,1,2,2,2,3,3)
  //C2 = (1,1,2,2,3,2,2,2,3,3,5,5)
  //Then, after making C1 and C2 disjoint, C1, C2 and I are:
  //C1 = (4,1)
  //C2 = (2,5,5)
  //I = {1,2,3}
  int i=0;
  while (i < C1.size())
  {
    int id = C1.get(i);
    if (C2.remove((Integer)id))
    {
      C1.remove(i);
      I.add(id);
    }
    else
     ++i;
  }

  //Decrement the internal reference count, irc, of the nodes
  //that have their ids in C2. If id is not in set I, add id
  //conditionally to set seedGarbageIds.
  for(Integer id2:C2)
  {
    decrIRC(id2);

    //If set I contains the id2 then p' will still refere to the
    //node having the id value id2.
    if (I.contains(id2))
      continue;

    //Node p' does not any more refer to a node that has the id
    //value id2. If we are not certain that the node is not
    //garbage add the id2 to the set seedGarbageIds if it is not
    //yet there.
    if (!isNodeCertainlyNotGarabge(id2))
      seedGarbageIds.add(id2);

  }

  //Increment the internal reference counts, irc, of the nodes
```

```
    //having their ids in C1.
    for(Integer id1:C1)
      incrIRC(id1);
  }


  //This method updates the object database with the white nodes
  //in a flat way. If a field of a white node p is a pointer field
  //then the id of a node in the field is copied to p', not the
  //node itself.
  private void copyContentsOfWhiteNodesToDB()
  throws Exception
  {
    for(Object p:whiteNodes)
      copyContentOfNodeToDB(p);
  }


  //This method updates the object database with the gray nodes
  //in a flat way. If a field of a gray node p is a pointer field
  //then the id of a node in the field is copied to p', not the
  //node itself. For simplicity, contents of all fields are
  //copied, ie. not only the changed fields. Also, to make
  //implementing easy, for a list node p the p' is first cleared
  //by making it an empty list.
  private void copyContentsOfGrayNodesToDB()
  throws Exception
  {
    for(Object p:grayNodes)
      copyContentOfGrayNodeToDB(p);
  }
  private void copyContentOfGrayNodeToDB(Object p)
  throws Exception
  {
    //This is a null operation for a fixed node.
    removeFieldsOfNodeInDB(p);

    copyContentOfNodeToDB(p);
  }


 //After the embed method has called the update method, it
 //calls the garbageCollection method to remove possible garbage
 //nodes from the object database.
 //
 //Real garbage nodes belong to the graph Z consisting of the
 //nodes reachable from nodes having their ids in the set
 //seedGarbageIds. Typically, only some or none of the nodes in Z
 //are garbage. Z may also be an empty set.
 //
 //The graph Z is examined by walking (traversing) it, each edge
 //in Z once. During walking, references for the reached nodes are
 //calculated. Walking produces for each node in Z the count of
 //incoming references in Z. By using this information and
 //reference count information (irc and orc) stored in the nodes
 //of Z it is determined which nodes in Z are referred to from
 //outside the Z. These nodes and reachable nodes from them are
 //not garbage. The remaining nodes in Z are real garbage.
 //
 //In some cases only part of the Z is needed to walk, i.e. Z can
 //be shrinkened.
 private void garbageCollection()
  throws Exception
  {
    //The map used to count incoming internal references in Z.
    countOfInternalReferencesInZ
      = new HashMap<Integer,Integer>();
    idsOfNodesRefOutsideZ = new HashSet<Integer>();

    //Walk the Z and calculate incoming internal references in Z.
    calculateReferencesProducedByWalkingInZ();
```

```
  //Determine the nodes in Z referred to from outside the Z.
  collectIdsOfNodesReferecedOutsideZ();

  //Determine non-garbage nodes in Z. Remaining nodes in Z are
  //real garbage nodes to be removed from the object database.
  removeIdsOfNonGarbageNodesInZ();

  //Remove garbage nodes from the database.
  removeGarbageNodesFromDB();

  //Free the global structures.
  countOfInternalReferencesInZ = null;
  idsOfNodesRefOutsideZ = null;
  seedGarbageIds = null;
}


//Calculate for each node in Z the count of incoming references
//in Z. For that we walk (traverse) the graph Z in the database,
//each edge once. The counts of incoming references in Z are
//collected in the map countOfInternalReferencesInZ where the
//key is the id of the node and the value is the count of the
//incoming references to that node in the Z.
private void calculateReferencesProducedByWalkingInZ()
throws Exception
{
  for (Integer seedGarbageId:seedGarbageIds)
  {
    //A trick:
    //Let p' be the node having id value seedGarbageId. If the
    //method call "walk(seedGarbageId)" walks to p' then the
    //count of incoming references for p' must be decreased by
    //one because the node p' is not reached through a real
    //edge in Z.
    if (walk(seedGarbageId))
      addToInternalReferencesInZ(seedGarbageId,-1);
  }
}
private boolean walk(Integer id)
throws Exception
{
  //Here we try to make the Z smaller, i.e. the node having "id"
  //is not walked to if we are sure that this node is not
  //garbage.
  if (isNodeCertainlyNotGarabge(id))
    return false;

  boolean nodeReachedBefore =
    countOfInternalReferencesInZ.containsKey(id);

  if (nodeReachedBefore) //The node has been seen before.
  {
    addToInternalReferencesInZ(id,1);
    return true;
  }

  //The node has not been seen before.
  countOfInternalReferencesInZ.put(id,1);

  //Walk to non-null child nodes.
  ArrayList<Integer> childIds =
    getIdsOfNonNullChildNodesOfDBNode(id);
  for(Integer idChild : childIds)
  {
    walk(idChild);
  }
  return true;
}
```

```
//This method returns true if we are sure that the node having
//the id is not garbage. However, in this demo implementation
//this method returns always false. However some chehckings
//could be done in real implementations. Some suggestions are in
//comments.
private boolean isNodeCertainlyNotGarabge(Integer id)
throws Exception
{
  //Possible checkings, for example:
  //1)
  //if "readORC(id) > 0" then the node is a persistent root node
  //and it cannot be garbage.
  //
  //2)
  //The node having the id can not be garbage if some node in the
  //node structure s has the same id and the root node s is a
  //white node. In this case the corresponding node s' is a
  //persistent root node and all reachable nodes from it are
  //persistent.
  //
  //3)
  //It is possible to implement an embed method which takes as a
  //parameter a list of ids of nodes which can not be garbage.
  //(The child program can know strategic nodes which are not
  //garbage) The parameter id of this method could be compared
  //to these ids.

  return false;
}


void addToInternalReferencesInZ(Integer id, int value)
{
  int oldValue = countOfInternalReferencesInZ.get(id);
  countOfInternalReferencesInZ.put(id,oldValue+value);
}


//Determine in the Z the nodes, their ids, which are referred to
//from outside the Z. These ids are collected in the set
//idsOfNodesRefOutsideZ.
private void collectIdsOfNodesReferecedOutsideZ()
throws Exception
{
  for (Map.Entry<Integer, Integer> e :
    countOfInternalReferencesInZ.entrySet())
  {
    int id = e.getKey();
    int countOfInternalReferences = e.getValue();
    int irc = readIRC(id);
    int orc = readORC(id);

    //Here we test if a node is referred to from outside the Z,
    //i.e. if a node is a persistent root node (orc >= 1) or
    //it is referred to from some node outside the Z
    //(countOfInternalReferences < irc). Note that always
    //countOfInternalReferences <= irc.
    //
    //Also note that if would filter (not done) in the method
    //isNodeCertainlyNotGarabge the nodes that have orc > 0 then
    //the test below ccould be replaced with the test
    //"if (countOfInternalReferences < irc)"
    if (countOfInternalReferences < orc + irc)
      idsOfNodesRefOutsideZ.add(id);
  }
}


//In the Z nodes reachable from nodes having they ids in the set
//idsOfNodesRefOutsideZ are not garbage. In this method ids of
//these nodes are removed from the map
```

```
  //countOfInternalReferencesInZ. The remaining nodes, having
  //their ids in the map countOfInternalReferencesInZ, are the
  //real garbage nodes.
  private void removeIdsOfNonGarbageNodesInZ()
  throws Exception
  {
    for(Integer id : idsOfNodesRefOutsideZ)
      removeIdOfNonGarbageNodeInZ(id);
  }
  private void removeIdOfNonGarbageNodeInZ(Integer id)
  throws Exception
  {
    if (countOfInternalReferencesInZ.remove(id) == null)
      return;

    ArrayList<Integer> childIds
      = getIdsOfNonNullChildNodesOfDBNode(id);
    for(Integer idChild : childIds)
      removeIdOfNonGarbageNodeInZ(idChild);
  }


  //Remove real garbage nodes from the database. These are the
  //nodes having their ids in the map
  //countOfInternalReferencesInZ.
  private void removeGarbageNodesFromDB()
  throws Exception
  {
    Set<Integer> keys = countOfInternalReferencesInZ.keySet();
    for(Integer id : keys)
      removeGarbageNodeFromDB(id);
  }
  private void removeGarbageNodeFromDB(int id)
  throws Exception
  {
    //Internal reference counts of (non-null) non-garbage
    //child nodes must be decremented accordingly.
    ArrayList<Integer> childIds =
      getIdsOfNonNullChildNodesOfDBNode(id);
    for(Integer idChild : childIds)
      if (!countOfInternalReferencesInZ.containsKey(idChild))
        decrIRC(idChild);

    Class<?> c = getClassOfDBNode(id);

    if (c != ListNode.class)
      removeFixedNodeFromDB(c,id);
    else
      removeListNodeFromDB(id);
  }
```

## 9. The Source Code of the Demo Implementation of the Embed Method

Below is full source code of the demo implementation of the embed method. It contains three files:

- TestDb.Java
- ListNode.java
- FieldT.java

### 9.1 File TestDB.java

```
/*
 * Copyright (c) 2016 Heikki Virkkunen.
```

```
 * Date: 5 April 2016
*/

package fi.heolvi.embed.base;

import java.lang.reflect.*;
import java.util.*;
import java.io.*;
import java.sql.*;

/*
This is the demo implementation of the embed method.

General information
-------------------
This implementation makes possible to test the embed method with
arbitrary complex update examples.

The implementation tries to be easy to read, rather than being
efficient as an implementation.

The SQLite database is used as an underlying data store.
Therefore, the implementation is formally an object-relational
(O/R) database, but the underlying data store could be, for
example, a random access file.

For simplicity, error checkings are not done. It is assumed that
a user does not give erroneous input for the embed method.

The term "node" is often used in the place of the term "object".

Nodes (objects) in the run-time memory are marked as s, p, q.
Their corresponding nodes in the object database are marked as s',
p', q' if they exist. However, we mark the nodes in the object
database as s', p' or q' even though the appropriate node s, p or
q does not exist in the run-time memory. The node s always denotes
a modified object structure (its root node) in the run-time memory
for which the embed method is called.


About the node types and the node ids
-------------------------------------
Each non-null node in the database has a unique id greater
than zero. The ids are not reused.

The implementation supports two kinds of nodes, ordinary nodes
(fixed nodes) and list nodes. Java types of fixed nodes (like
Person, Friend etc.) are defined by the client program. Each
type of the fixed node has its own SQL table created by the
system automatically when the (Java) type is encountered for the
first time.

For simplicity, inheritance is not supported for the Java types
of the nodes.

Fields of the nodes are either of type scalar or pointer.

For simplicity, the allowed scalar types are only the Java types
int, Integer and String. In list nodes the type int is replaced
automatically with the type Integer, because a run-time list node
stores its items into a Java array list.

In the run-time memory a pointer field of a node contains
reference to the referred node (Java object). In the database
a pointer field contains the id of the referred node, or zero
if the field refers to a null node. Null nodes are not stored
explicitly in the database.

When a node p' is allocated in the object database, it gets
an id>0 which has not been used before in the database. At the
same time the same id is bound to p, i.e. stored in its id field.
```

```
In the run-time memory, the type of the field of a fixed node is
determined from the field but the type of the field of a list
node is determined from the content of the field. Single list
node can contain items of different types, fore example scalar and
pointer values. To avoid ambiguity, a null value in a run-time
list node is interpreted as a special value "None". In the
database, fields of the list nodes have explicit types.

Each run-time node has a compulsory id field of type int storing
the id of corresponding node in the object database if that
exists. If a node p in the run-time memory has id>0, it has a
corresponding node p' in the database having that id. Otherwise,
when the id is zero for p, then p does not have a corresponding
node in the database.
*/




public class TestDB
{

  //Name of the database.
  private String dbName;

  //Java JDBC connection of the database.
  private Connection connection;

  //Id of a null node is zero in the run-time memory and in the
  //database.
  public static int ID_NULL_NODE = 0;


  //Id value used for a non-null run-time node p if p does not
  //yet have a corresponding node p' in the database.
  public static int ID_ZERO = 0;

  //In the run-time memory, a null value in a list node
  //requires special handling because a null value does
  //not have explicit type in Java. Therefore, in this context,
  //the type of the null is interpreted as a special type "NONE".
  public static int FIELD_TYPE_NONE              = 0;

  //Here are the supported scalar types for the fields of the
  //nodes.
  //Java type int is supported only for fixed nodes.
  public static int FIELD_TYPE_INT     = 1;
  //Java type Integer is supported for fixed nodes and list nodes.
  public static int FIELD_TYPE_INTEGER = 2;
  //Java type String is supported for fixed nodes and list nodes.
  public static int FIELD_TYPE_STRING  = 3;

  //Supported pointer fields for the nodes are pointer to a fixed
  //node and pointer to a list node.
  public static int FIELD_TYPE_FIXED_NODE        = 4;
  public static int FIELD_TYPE_LIST_NODE         = 5;

  //The embed method is called for a modified object structure s
  //(with its root node) in the run-time memory. In the beginning
  //of the embed method non-null nodes in s are separated in two
  //groups, white nodes and gray nodes. A non-null node p in s is
  //a white node if its id is zero, i.e. if p does not yet have a
  //corresponding node in the the database. Otherwise, if id>0, p
  //already has a corresponding node in the database with that id
  //and p is a gray node. White nodes are collected in the list
  //whiteNodes and gray nodes are collected in the list GrayNodes.
  ArrayList<Object> whiteNodes;
  ArrayList<Object> grayNodes;

  //As a side effect, the update phase of the embed method
```

```
  //collects ids of potential garbage nodes, it finds, in the
  //set seedGarbageIds. When the update phase has been finished,
  //all possible garbage nodes belong to set or graph Z
  //consisting of the nodes reachable from the nodes having their
  //ids in seedGarbageIds. It is possible that only some, or none
  //of the nodes in Z are garbage. Z may also be empty.
  HashSet<Integer> seedGarbageIds;

  //The map countOfInternalReferencesInZ is used to count the
  //internal references inside Z. The key is id of a node and the
  //value is count of internal references to that node in Z.
  HashMap<Integer,Integer> countOfInternalReferencesInZ;

  //In Z, nodes which are referred to from outside the Z,
  //cannot be garbage. Ids of those nodes are collected in the
  //set idsOfNodesRefOutsideZ.
  HashSet<Integer> idsOfNodesRefOutsideZ;

  //This map is for a tool method search which can be used to load
  //(search) object structures from the object database into the
  //run-time memory.
  HashMap<Integer,Object> readNodes;



/////////////////////////////////////////////////////////////////
// Public methods.

  //This method is constructor of the object database.
  //The client program calls it to create a new object database or
  //to open an existing database. The method call is like:
  //
  //  TestDB db = new TestDB(dbName);
  //
  //The method creates a new database if the database with a given
  //name does not exist yet. When the database is created, initial
  //system tables for it are created at the same time. This
  //includes creating two tables to store list nodes.
  //
  //Tables for different types of fixed nodes are created
  //dynamically on-fly in the method
  //methodcreateDBTableForFixedNodewhen when the appropriate type
  //is encountered for the first time.
  public TestDB(String dbName)
  throws Exception
  {
    this.dbName = dbName;

    boolean isOldDB = existsDB(dbName);

    try
    {
      Class.forName("org.sqlite.JDBC");
      connection =
        DriverManager.getConnection("jdbc:sqlite:"+dbName);
    }
    catch ( Exception e )
    {
      System.out.println( e.getClass().getName() +
        ": " + e.getMessage() );
      System.exit(0);
    }


    if (isOldDB)
      return;


    //Higher level structure of a node instance is stored in the
    //nodeInstances table.
    String str =
```

```
  "CREATE TABLE nodeInstances"
  +" ("

  //Unique id of the node in the object database.
  +"id INTEGER PRIMARY KEY AUTOINCREMENT,"

  //Count of outer references (orc) to the node.
  +"orc INTEGER,"

  //Count of internal references (irc) to the node.
  +"irc INTEGER,"

  //Full Java type name of the node.
  //For a list node the name is "testdb.ListNode"
  //For a fixed node the type name is, for example,
  //"userclasses.Friend".
  //(It is inefficient to store for each node instance the
  //full Java type name, but this is a demo implementation.)
  +"className TEXT"

  +")";
executeStatement(str);


//List node instances are stored in the tables "list" and
//"listitems" in th database.

str =
  "CREATE TABLE lists"
  +" ("

  //Primary key. The listItems table refers to this field.
  +"id INTEGER PRIMARY KEY AUTOINCREMENT,"

  //Refers to nodeInstances.id.
  +"instanceId INTEGER,"

  //Length of the list node, i.e. how many items the list node
  //contains currently.
  +"len INTEGER"

  +")";
executeStatement(str);


str =
  "CREATE TABLE listItems"
  +" ("

  //Primary key.
  +"id INTEGER PRIMARY KEY AUTOINCREMENT,"

  //Refers to lists.id.
  +"parent INTEGER,"

  //Position (0,1,..) of an item in the list.
  +"position INTEGER,"

  //Type code of an item in a list.
  //Defines explicitly the type of the item.
  +"type INTEGER,"

  //An item in a list. The item can be a scalar value or
  //pointer to node, to a fixed node or to a list node. If
  //an item is a scalar value, it is stored as such. Note that
  //SQLite database uses dynamic type system. Therefore also
  //a string scalar can be stored in this field. If item is
  //pointer to a node then item (integer value) refers to
  //nodeInstances.id.
  +"item INTEGER"
```

```
    +")";

  executeStatement(str);
}


//This method closes the database.
public void close()
throws Exception
{
  connection.close();
}


//This method tests whether the database with a given name
//exists.
public static boolean existsDB(String dbName)
throws Exception
{
  return (new File(dbName)).exists();
}


//This method deletes the database with a given name.
public static void deleteDB(String dbName)
throws Exception
{
  (new File(dbName)).delete();
}


//This is the famous embed method which a client program calls to
//update the object database with a modified object structure s
//in the run-time memory. The method is called with the root
//object s of the modified object structure. The embed method
//consists of two co-operating phases, the update phase and the
//garbage collection phase.
public void embed(Object s)
throws Exception
{
  update(s);
  garbageCollection();
}


//This is a tool method. Method incrORC can be called for a
//non-null run-time node p having a corresponding node p' in the
//object database. Method incrORC increments the outer reference
//count, orc, of p' by one. Note that it is allowed that orc of
//p' gets greater values than one. If orc>1 it can be thought
//that a node p' has been saved several times from outside the
//database or that many users refer to it from outside the
//database.
public void incrORC(Object p)
throws Exception
{
  incrORC(getId(p));
}


//This is a tool method. The method can be understood as a
//delete operation for a node structure p' in the object
//database. Method decrORC can be called for a non-null run-time
//node p having a corresponding node p' in the object database.
//It is assumed that the outer reference count, orc of p' is
//greater than zero before calling the method. Error checking is
//not done. Method decrORC decrements the orc of p' by one and
//calls garbage collection for p' because now p' and some nodes
//reachable from it can be garbage.
public void decrORC(Object p)
throws Exception
```

```
  {
    //Create the global set seedGarbageIds.
    seedGarbageIds = new HashSet<Integer>();

    decrORC(getId(p));

    seedGarbageIds.add(getId(p));

    garbageCollection();
  }


  //This is a tool method. A client program can use this method to
  //load (search) object structures from the object database back
  //to the run-time memory.
  //
  //Example invocation:
  //
  //db.searchFixedNodesFromDB(Friend.class,"name","A",
  //                          Friend.class,"age","23",
  //                          Book.class,"age","23"
  //                          )
  //
  //This invocation performs three searches and returns
  //ArrayList<Object> of three items, having type Friend, Friend
  //and Book, respectively. Also reachable nodes from these nodes
  //are returned, i.e. loading is a deep operation. If the same
  //node instance exists several times in the returned node
  //structures, it is returned only once.
  //
  //If a search condition match for several nodes, only the
  //first node that was found, and its reachable nodes, are
  //returned.
  //
  //It is possible that a search condition does not match
  //for any node.
  //
  //The search condition contains three items:
  //  The type of the node (for example Friend.class).
  //  The name of the scalar field (for example "name").
  //  The scalar value (compatible with the type of the search
  //  field) used to test equality in a search operation
  //(for example "A").
  public ArrayList<Object> searchFixedNodesFromDB
    (Object... searchRules)
  throws Exception
  {
    ArrayList<Object> resultNodes = new ArrayList<Object>();

    readNodes = new HashMap<Integer,Object>();
    int rootCount = searchRules.length/3;
    for (int i=0; i<rootCount; ++i)
    {
      int j = 3*i;
      Class<?> classOfFixedNode     = (Class<?>) searchRules[j];
      String fieldNameOfScalarField = (String)   searchRules[j+1];
      Object value                  =            searchRules[j+2];

      Object node = searchFixedNodeFromDB(classOfFixedNode,
                      fieldNameOfScalarField, value);
      if (node != null)
        resultNodes.add(node);
    }
    return resultNodes;
  }
// Public methods.
//////////////////////////////////////////////////////////////////



//////////////////////////////////////////////////////////////////
```

```
// Higher level private methods.

  //The embed method calls the update method with the root node s
  //of the modified object data structure in the run-time memory.
  //After the update method has been finished, content of the
  //object structure s exists in the database.
  //
  //As a side effect, the update method produces the set of ids of
  //potential garbage nodes, it finds, into the set seedGarbageIds.
  //The set seedGarbageIds is input for the garbage collection
  //phase of the embed method.
  //
  //Below the terms "step 1", "step 2", "step 3" and "step 4"
  //refer to the steps described in the article, in section 4.
  private void update(Object s)
  throws Exception
  {
    //Global lists for the white nodes and the gray nodes in s.
    whiteNodes = new ArrayList<Object>();
    grayNodes = new ArrayList<Object>();

    //This boolean value dscribes whether the root node s is
    //white. This information is used in the step 4 of this
    //method.
    boolean rootIsWhite =
      getId(s) == ID_ZERO;

    //Create the global set seedGarbageIds.
    seedGarbageIds = new HashSet<Integer>();

//Step 1 and step 2 of the update method:
    //Step1: Collect white nodes in object the structure s in the
    //list whiteNodes and gray nodes in s in the list grayNodes.
    //Step 2: For each white node in s: Allocate the same type of
    //empty node in the database.
    collectWhiteAndGrayNodes(s);

//Step 3 of the update method:
    //Handle changes of internal reference counts of the nodes
    //caused by updating the database with the white nodes.
    handleReferencesFromWhiteNodesInDB();

    //Handle changes of internal reference counts of the nodes
    //caused by updating the database with the gray nodes.
    handleReferencesFromGrayNodesInDB();

    //Update the database with the white nodes.
    copyContentsOfWhiteNodesToDB();

    //Update the database with the gray nodes.
    copyContentsOfGrayNodesToDB();


//Step 4 of the update method:
    //For selected white nodes in s make their corresponding nodes
    //in the database persistent root nodes by setting them orc=1.
    //
    //We simply select only the root node s, if it is a white node.
    //Otherwise no nodes are selected.
    //
    //When a root node s is a white node it is obvious that the
    //user wants to make s' a persistent root node to make s' and
    //all the reachable nodes from it persistent in the database.
    //
    //There exist rare cases where a user may want that for a
    //white root node s the corresponding node s' gets orc=0 and
    //does not not become a persistent root node in the database.
    //For example if the run-time object structure s is a circular
    //structure s -> p -> s, where s is a white node and p is a
    //gray node, a user may want that persistence of s' depends on
    //the persistence of p'. Therefore orc could be zero for s in
```

```
  //this case.
  //
  //If the embed method is called for a gray node s, then the
  //orc of s' is kept as it was. This is a natural decision.
  //However, this can cause a strange but perhaps correct
  //effects in some cases; the node s', and perhaps some
  //reachable nodes from it can disappear from the object
  //database as a consequence of an executed embed method! The
  //following example describes this.
  //
  //Let s' originally refer to some node p' and p' refer back to
  //s' and suppose that no-one else in the database is referring
  //to s', and that orc=0 for s'. After that circular structure
  //s -> p -> s is modified in the run-time memory by setting p
  //to refer to a null node, i.e. not to s anymore. After that
  //the embed method is called for s. As a consequence s'
  //becomes garbage, because now irc=orc=0 for s'.
  if (rootIsWhite)
    incrORC(getId(s));

  //Free the lists reserved for the white nodes and the gray nodes.
  whiteNodes = null;
  grayNodes = null;
}


//Method collectWhiteAndGrayNodes is called from the method
//update. This method separates white nodes and gray nodes in the
//object structure s. White nodes are collected in the list
//whiteNodes and gray nodes in the list grayNodes.
//
//In addition, the method allocates for each white node p in s a
//corresponding empty node p' (having a corresponding type) in
//the object database. The node p gets the id of p' into its
//id field.
private void collectWhiteAndGrayNodes(Object p)
throws Exception
{
  //Null nodes are note collected.
  if (p==null)
    return;

  //The same node instance is not collected twice.
  if (bufferContainsNode(whiteNodes,p)
      || bufferContainsNode(grayNodes,p))
    return;

  int id = getId(p);

  //If p is a white node, collect p and allocate a corresponding empty
  //node p' in the database. Assign id of p' to p.
  if (id == ID_ZERO)
  {
    id = allocateNodeInDB(p);
    setId(p,id);
    whiteNodes.add(p);
  }
  else //Collect a gray node p.
    grayNodes.add(p);

  //Collect non-null child nodes of p if not yet collected.
  ArrayList<FieldT> pointerFields = getFields(p.getClass(),p,1);
  for (FieldT f : pointerFields)
    collectWhiteAndGrayNodes(f.value);
}


//Handle changes of internal reference counts caused by updating
//the object database with white nodes.
private void handleReferencesFromWhiteNodesInDB()
throws Exception
```

```
{
  for(Object p:whiteNodes)
    handleReferencesFromWhiteNodeInDB(p);
}
//Here p is a white node. Updating p' with the p can increase
//internal reference counts of some nodes in the database. These
//changes are updated in this method.
//
//The following rule gives the result:
//If a white node p refers with a pointer field f to a non-null
//node q then the internal reference count of the node q'
//increments by one. If p refers to q many times (with many
//pointer fields), then the internal reference count of q'
//increments as many times.
private void handleReferencesFromWhiteNodeInDB(Object p)
throws Exception
{
  ArrayList<Integer> C1 = getIdsOfNonNullChildNodes(p);
  for(Integer id:C1)
    incrIRC(id);
}


//Handle changes of internal reference counts caused by updating
//the object database with gray nodes.
private void handleReferencesFromGrayNodesInDB()
throws Exception
{
  for(Object p:grayNodes)
    handleReferencesFromGrayNodeInDB(p);
}
//Here p is a gray node. Updating p' with the p can change
//internal reference counts of some nodes in the database. These
//changes are updated in this method.
//
//The following rules give the result:
//
//1)
//If, before updating the node p' with a gray node p, the node
//p' refers with a pointer field f' to a non-null node q' then
//the internal reference count of q' decrements by one. If p'
//refers to q' many times (with many pointer fields), then the
//internal reference count of q' decrements as many times.
//
//2)
//If a gray node p refers with a pointer field f to a non-null
//node q then the internal reference count of the node q'
//increments by one. If p refers to q many times (with many
//pointer fields), then the internal reference count of q'
//increments as many times.
//
//The algorithm does incrementing/decrementing in such a way
//that the internal reference count of a node is only
//incremented or decremented, not both.
//
//If, after updating, the node p' does not any more refer to a
//non-null node q', then q' may be garbage. In this case the id
//of q' is added conditionally to set seedGarbageIds, if it is
//not yet there.
private void handleReferencesFromGrayNodeInDB(Object p)
throws Exception
{

  //Construct the list C1 = (id(q1),..,id(qn)) of the ids of
  //non-null child nodes of the node p. If p refers several
  //times to the same non-null child node q then the id of q is
  //as many times in the list C1.
  ArrayList<Integer> C1 =
    getIdsOfNonNullChildNodes(p);
```

```
  //Construct the list C2 = (id(q'1),..,id(q'm)) of the ids of
  //non-null child nodes of the node p' (before p' has been
  //updated with p). If p' refers several times to the same
  //non-null child node q' then the id of q' is as many times
  //in the list C2.
  ArrayList<Integer> C2 =
    getIdsOfNonNullChildNodesOfDBNode(getId(p));

  //The set I will contain the intersection of C1 and C2. (The
  //same id is not twice in I).
  HashSet<Integer> I = new HashSet<Integer>();

  //Make lists C1 and C2 disjoint. The intersection of C1 and C2
  //is collected in the set I. Note that C1 can contain the
  //same id several times, before and after making C1 and C2
  //disjoint. The same is true for the list C2.
  //For example: Let
  //C1 = (1,2,1,3,4,1,2,2,2,3,3)
  //C2 = (1,1,2,2,3,2,2,2,3,3,5,5)
  //Then, after making C1 and C2 disjoint, C1, C2 and I are:
  //C1 = (4,1)
  //C2 = (2,5,5)
  //I = {1,2,3}
  int i=0;
  while (i < C1.size())
  {
    int id = C1.get(i);
    if (C2.remove((Integer)id))
    {
      C1.remove(i);
      I.add(id);
    }
    else
     ++i;
  }

  //Decrement the internal reference count, irc, of the nodes
  //that have their ids in C2. If id is not in set I, add id
  //conditionally to set seedGarbageIds.
  for(Integer id2:C2)
  {
    decrIRC(id2);

    //If set I contains the id2 then p' will still refere to the
    //node having the id value id2.
    if (I.contains(id2))
      continue;

    //Node p' does not any more refer to a node that has the id
    //value id2. If we are not certain that the node is not
    //garbage add the id2 to the set seedGarbageIds if it is not
    //yet there.
    if (!isNodeCertainlyNotGarabge(id2))
      seedGarbageIds.add(id2);

  }

  //Increment the internal reference counts, irc, of the nodes
  //having their ids in C1.
  for(Integer id1:C1)
    incrIRC(id1);
}


//This method updates the object database with the white nodes
//in a flat way. If a field of a white node p is a pointer field
//then the id of a node in the field is copied to p', not the
//node itself.
private void copyContentsOfWhiteNodesToDB()
throws Exception
{
```

```
    for(Object p:whiteNodes)
      copyContentOfNodeToDB(p);
  }


  //This method updates the object database with the gray nodes
  //in a flat way. If a field of a gray node p is a pointer field
  //then the id of a node in the field is copied to p', not the
  //node itself. For simplicity, contents of all fields are
  //copied, i.e. not only the changed fields. Also, to make
  //implementing easy, for a list node p the p' is first cleared
  //by making it an empty list.
  private void copyContentsOfGrayNodesToDB()
  throws Exception
  {
    for(Object p:grayNodes)
      copyContentOfGrayNodeToDB(p);
  }
  private void copyContentOfGrayNodeToDB(Object p)
  throws Exception
  {
    //This is a null operation for a fixed node.
    removeFieldsOfNodeInDB(p);

    copyContentOfNodeToDB(p);
  }


 //After the embed method has called the update method, it
 //calls the garbageCollection method to remove possible garbage
 //nodes from the object database.
 //
 //Real garbage nodes belong to the graph Z consisting of the
 //nodes reachable from nodes having their ids in the set
 //seedGarbageIds. Typically, only some or none of the nodes in Z
 //are garbage. Z may also be an empty set.
 //
 //The graph Z is examined by walking (traversing) it, each edge
 //in Z once. During walking, references for the reached nodes are
 //calculated. Walking produces for each node in Z the count of
 //incoming references in Z. By using this information and
 //reference count information (irc and orc) stored in the nodes
 //of Z it is determined which nodes in Z are referred to from
 //outside the Z. These nodes and reachable nodes from them are
 //not garbage. The remaining nodes in Z are real garbage.
 //
 //In some cases only part of the Z is needed to walk, i.e. Z can
 //be shrunk.
 private void garbageCollection()
  throws Exception
  {
    //The map used to count incoming internal references in Z.
    countOfInternalReferencesInZ
      = new HashMap<Integer,Integer>();
    idsOfNodesRefOutsideZ = new HashSet<Integer>();

    //Walk the Z and calculate incoming internal references in Z.
    calculateReferencesProducedByWalkingInZ();

    //Determine the nodes in Z referred to from outside the Z.
    collectIdsOfNodesReferecedOutsideZ();

    //Determine non-garbage nodes in Z. Remaining nodes in Z are
    //real garbage nodes to be removed from the object database.
    removeIdsOfNonGarbageNodesInZ();

    //Remove garbage nodes from the database.
    removeGarbageNodesFromDB();

    //Free the global structures.
    countOfInternalReferencesInZ = null;
```

```
  idsOfNodesRefOutsideZ = null;
  seedGarbageIds = null;
}


//Calculate for each node in Z the count of incoming references
//in Z. For that we walk (traverse) the graph Z in the database,
//each edge once. The counts of incoming references in Z are
//collected in the map countOfInternalReferencesInZ where the
//key is the id of the node and the value is the count of the
//incoming references to that node in the Z.
private void calculateReferencesProducedByWalkingInZ()
throws Exception
{
  for (Integer seedGarbageId:seedGarbageIds)
  {
    //A trick:
    //Let p' be the node having id value seedGarbageId. If the
    //method call "walk(seedGarbageId)" walks to p' then the
    //count of incoming references for p' must be decreased
    //afterward by one because the node p' is not reached through
    //a real edge in Z.
    if (walk(seedGarbageId))
      addToInternalReferencesInZ(seedGarbageId,-1);
  }
}
private boolean walk(Integer id)
throws Exception
{
  //Here we try to make the Z smaller, i.e. to the node having
  //"id" is not walked to if we are sure that this node is not
  //garbage.
  if (isNodeCertainlyNotGarabge(id))
    return false;

  boolean nodeReachedBefore =
    countOfInternalReferencesInZ.containsKey(id);

  if (nodeReachedBefore) //The node has been seen before.
  {
    addToInternalReferencesInZ(id,1);
    return true;
  }

  //The node has not been seen before.
  countOfInternalReferencesInZ.put(id,1);

  //Walk to non-null child nodes.
  ArrayList<Integer> childIds =
    getIdsOfNonNullChildNodesOfDBNode(id);
  for(Integer idChild : childIds)
  {
    walk(idChild);
  }
  return true;
}


//This method returns true if we are sure that the node having
//the id is not garbage. However, in this demo implementation
//this method returns always false. Some checkings
//could be done in real implementations. Some suggestions are in
//comments.
private boolean isNodeCertainlyNotGarabge(Integer id)
throws Exception
{
  //Possible checkings, for example:
  //1)
  //if "readORC(id) > 0" then the node is a persistent root node
  //and it cannot be garbage.
  //
```

```
  //2)
  //The node having the id can not be garbage if some node in the
  //node structure s has the same id and the root node s is a
  //white node. In this case the corresponding node s' is a
  //persistent root node and all reachable nodes from it are
  //persistent.
  //
  //3)
  //It is possible to implement an embed method which takes as a
  //parameter a list of ids of nodes which can not be garbage.
  //(The child program can know strategic nodes which are not
  //garbage) The parameter id of this method could be compared
  //to these ids.

  return false;
}


void addToInternalReferencesInZ(Integer id, int value)
{
  int oldValue = countOfInternalReferencesInZ.get(id);
  countOfInternalReferencesInZ.put(id,oldValue+value);
}


//Determine in the Z the nodes, their ids, which are referred to
//from outside the Z. These ids are collected in the set
//idsOfNodesRefOutsideZ.
private void collectIdsOfNodesReferecedOutsideZ()
throws Exception
{
  for (Map.Entry<Integer, Integer> e :
    countOfInternalReferencesInZ.entrySet())
  {
    int id = e.getKey();
    int countOfInternalReferences = e.getValue();
    int irc = readIRC(id);
    int orc = readORC(id);

    //Here we test if a node is referred to from outside the Z,
    //i.e. if a node is a persistent root node (orc >= 1) or
    //it is referred to from some node outside the Z
    //(countOfInternalReferences < irc). Note that always
    //countOfInternalReferences <= irc.
    //
    //Also note that if would filter (not done) in the method
    //isNodeCertainlyNotGarabge the nodes that have orc > 0 then
    //the test below ccould be replaced with the test
    //"if (countOfInternalReferences < irc)"
    if (countOfInternalReferences < orc + irc)
      idsOfNodesRefOutsideZ.add(id);
  }
}


//In the Z nodes reachable from nodes having they ids in the set
//idsOfNodesRefOutsideZ are not garbage. In this method ids of
//these nodes are removed from the map
//countOfInternalReferencesInZ. The remaining nodes, having
//their ids in the map countOfInternalReferencesInZ, are the
//real garbage nodes.
private void removeIdsOfNonGarbageNodesInZ()
throws Exception
{
  for(Integer id : idsOfNodesRefOutsideZ)
    removeIdOfNonGarbageNodeInZ(id);
}
private void removeIdOfNonGarbageNodeInZ(Integer id)
throws Exception
{
  if (countOfInternalReferencesInZ.remove(id) == null)
```

```
      return;

    ArrayList<Integer> childIds
      = getIdsOfNonNullChildNodesOfDBNode(id);
    for(Integer idChild : childIds)
      removeIdOfNonGarbageNodeInZ(idChild);
  }


  //Remove real garbage nodes from the database. These are the
  //nodes having their ids in the map
  //countOfInternalReferencesInZ.
  private void removeGarbageNodesFromDB()
  throws Exception
  {
    Set<Integer> keys = countOfInternalReferencesInZ.keySet();
    for(Integer id : keys)
      removeGarbageNodeFromDB(id);
  }
  private void removeGarbageNodeFromDB(int id)
  throws Exception
  {
    //Internal reference counts of (non-null) non-garbage
    //child nodes must be decremented accordingly.
    ArrayList<Integer> childIds =
      getIdsOfNonNullChildNodesOfDBNode(id);
    for(Integer idChild : childIds)
      if (!countOfInternalReferencesInZ.containsKey(idChild))
        decrIRC(idChild);

    Class<?> c = getClassOfDBNode(id);

    if (c != ListNode.class)
      removeFixedNodeFromDB(c,id);
    else
      removeListNodeFromDB(id);
  }


  //Remove a fixed node from the database, in a flat way.
  private void removeFixedNodeFromDB(Class<?> c, int id)
  throws Exception
  {
    String tableName =
      getFixedTableNameFromClassName(c.getName());
    executeDelete(tableName, "instanceId=?", id);
    executeDelete("nodeInstances", "id=?", id);
  }


  //Remove a list node from the database, in a flat way.
  private void removeListNodeFromDB(int id)
  throws Exception
  {
    int rowIdOfList = (Integer)
      readSingleValue("lists","id","instanceId=?",id);
    executeDelete("listItems", "parent=?", rowIdOfList);
    executeDelete("lists", "id=?", rowIdOfList);
    executeDelete("nodeInstances", "id=?", id);
  }
// Higher level private methods.
//////////////////////////////////////////////////////////////////


//////////////////////////////////////////////////////////////////
// Tool methods for the run-time memory nodes.
  private int getId(Object p)
  throws Exception
  {
    if (p==null)
      return ID_NULL_NODE;
```

```
  Class c = p.getClass();
  Field f = c.getDeclaredField("id");
  f.setAccessible(true);
  return (Integer) f.get(p);
}


private void setId(Object p, int id)
throws Exception
{
  Class<?> c = p.getClass();
  Field f = c.getDeclaredField("id");
  f.setAccessible(true);
  f.set(p,id);
}


private ArrayList<Integer> getIdsOfNonNullChildNodes(Object p)
throws Exception
{
   ArrayList<Integer> ids = new ArrayList<Integer>();
   ArrayList<FieldT> fields = getFields(p.getClass(),p,1);
   for(FieldT f: fields)
   {
     int id = getId(f.value);
     if (id != ID_NULL_NODE)
       ids.add(id);
   }
   return ids;
}


//Fields of the run-time node p are returned in a ArrayList.
//Both a field and its content is returned in the structure
//FieldT. A caller selects with the selector if only scalar
//fields or only pointer fields, or if both types of fields
//are returned.
//
//The c is class of node (fixed or list node).
//If method is called for a fixed node,
//the null p may be given as a parameter.
//In this case no values are returned in the fields, i.e.
//FieldT.value is not filled in this case.
//Values of the selector:
//0 = Pointer fields and scalar fields are returned.
//1 = Only pointer fields are returned.
//2 = Only scalar fields are returned.
private ArrayList<FieldT>
  getFields(Class<?> c, Object p, int selector)
throws Exception
{
  if (c != ListNode.class)
    return getFieldsOfFixedNode(c,p,selector);
  else
    return getFieldsOfListNode((ListNode)p,selector);
}
private ArrayList<FieldT>
  getFieldsOfFixedNode(Class<?> c, Object p, int selector)
throws Exception
{
  ArrayList<FieldT> fields = new ArrayList<FieldT>();
  Field[] allFields = c.getDeclaredFields();
  for(Field f: allFields)
  {
    if (f.getName().equals("id"))
      continue;

    Class<?> cf = f.getType();
    int typeCodeOfField = getTypeCodeOfFieldFromClass(cf);
```

```
      boolean addPointerField =
        isPointerField(typeCodeOfField) &&
          (selector == 0 || selector == 1);
      boolean addScalarField =
        isScalarField(typeCodeOfField) &&
          (selector == 0 || selector == 2);

      if (addPointerField || addScalarField)
      {
        Object value = null;
        if (p != null)
        {
          f.setAccessible(true);
          value = f.get(p);
        }
        fields.add(new FieldT(f,typeCodeOfField,value));
      }
    }
    return fields;
  }


  private ArrayList<FieldT>
    getFieldsOfListNode(ListNode p, int selector)
  throws Exception
  {
    ArrayList<FieldT> fields = new ArrayList<FieldT>();

    int len = p.list.size();
    for(int pos=0;pos<len;++pos)
    {
      Object value = p.list.get(pos);
      int typeCodeOfField =
        getTypeCodeOfFieldFromValueInField(value);

      boolean addPointerField =
        isPointerField(typeCodeOfField) &&
          (selector == 0 || selector == 1);
      boolean addScalarField =
        isScalarField(typeCodeOfField) &&
          (selector == 0 || selector == 2);

      if (addPointerField || addScalarField)
        fields.add(new FieldT(pos,typeCodeOfField,value));

    }//for
    return fields;
  }


  private boolean
    bufferContainsNode(ArrayList<Object> buffer, Object p)
  {
    for (Object o:buffer)
      if (o == p) return true;
    return false;
  }


  private boolean grayNodesContainsId(int id)
  throws Exception
  {
    for(Object p:grayNodes)
      if (getId(p) == id)
        return true;
    return false;
  }
// Tool methods for the run-time memory nodes.
/////////////////////////////////////////////////////////////////
```

```
////////////////////////////////////////////////////////////////////
// Tool methods for the nodes in DB.
  private int allocateNodeInDB(Object p)
  throws Exception
  {
    Class<?> c = p.getClass();
    if (c != ListNode.class)
      return allocateFixedNodeInDB(p);
    else
      return allocateListNodeInDB();

  }
  private int allocateFixedNodeInDB(Object p)
  throws Exception
  {
    Class<?> c = p.getClass();
    String tableName =
      getFixedTableNameFromClassName(c.getName());

    //If a node type is a new one, create the corresponding SQL
    //table.
    if (!tableExists(tableName))
      createDBTableForFixedNode(c);

    //Insert an empty fixed node. Set orc=irc=0 for the inserted
    //node.
    int id = doInsertReturnPrimaryKey
      ("INSERT INTO nodeInstances (id,orc,irc,className)
VALUES(NULL,0,0,?)",c.getName());
    executeStatement
      ("INSERT INTO " + tableName + " (id,instanceId) VALUES(NULL,?)",id);

    return id;
  }
  private int allocateListNodeInDB()
  throws Exception
  {
   //Insert an empty list node. Set orc=irc=0 for the inserted
   //node.
   int id = doInsertReturnPrimaryKey
     ("INSERT INTO nodeInstances (id,orc,irc,className) VALUES(NULL,0,0,?)",
      ListNode.class.getName());

   executeStatement("INSERT INTO lists VALUES(NULL,?,0)",id);

   return id;
  }


  private void removeFieldsOfNodeInDB(Object p)
  throws Exception
  {
    Class<?> c = p.getClass();
    if (c != ListNode.class)
      ;
    else
      removeFieldsOfListNodeInDB(getId(p));
  }
  private void removeFieldsOfListNodeInDB(int id)
  throws Exception
  {
    int rowIdOfList = (Integer) readSingleValue
      ("lists","id","instanceId=?",id);

    executeDelete("listItems", "parent=?", rowIdOfList);

    executeStatement
      ("UPDATE lists SET len=0 WHERE id=?", rowIdOfList);
  }
```

```
private void copyContentOfNodeToDB(Object p)
throws Exception
{
  Class<?> c = p.getClass();
  if (c != ListNode.class)
    copyContentsOfFixedNodeToDB(p);
  else
    copyContentsOfListNodeToDB((ListNode)p);
}
private void copyContentsOfFixedNodeToDB(Object p)
throws Exception
{
  ArrayList<FieldT> fields = getFields(p.getClass(),p,0);
  for (FieldT f : fields)
    writeValueToFieldOfDBNode(p,f);
}
private void copyContentsOfListNodeToDB(ListNode p)
throws Exception
{
  ArrayList<FieldT> fields = getFields(p.getClass(),p,0);
  for (FieldT f : fields)
    writeValueToFieldOfDBListNode(p,f);
}


private void writeValueToFieldOfDBNode(Object p, FieldT f)
throws Exception
{
  Class<?> c = p.getClass();
  if (c != ListNode.class)
    writeValueToFieldOfDBFixedNode(p,f);
  else
    writeValueToFieldOfDBListNode(p,f);
}
private void writeValueToFieldOfDBFixedNode(Object p, FieldT f)
throws Exception
{
  Object valueDB = f.value;

  if (isPointerField(f.typeCode))
    valueDB = getId(valueDB);

  Class<?> c = p.getClass();
  String tableName =
    getFixedTableNameFromClassName(c.getName());

  int rowId = (Integer) readSingleValue
    (tableName,"id","instanceId=?",getId(p));

  String fieldName = ((Field) f.field).getName();

  updateSingleValue(tableName,fieldName,"id=?",valueDB,rowId);
}
private void writeValueToFieldOfDBListNode(Object p, FieldT f)
throws Exception
{
  Object valueDB = f.value;

  if (isPointerField(f.typeCode))
    valueDB = getId(valueDB);

  int rowIdOfList = (Integer) readSingleValue
    ("lists","id","instanceId=?",getId(p));

  executeStatement("INSERT INTO listItems VALUES(NULL,?,?,?,?)",
                   rowIdOfList,
                   (Integer)f.field,
                   f.typeCode,
                   valueDB);

  executeStatement
```

```
    ("UPDATE lists SET len=len+1 WHERE id=?",rowIdOfList);
}


private ArrayList<Integer>
  getIdsOfNonNullChildNodesOfDBNode(int id)
throws Exception
{

   ArrayList<Integer> ids = new ArrayList<Integer>();
   ArrayList<FieldT> fields = readFieldsOfDBNode(id,1);
   for(FieldT f: fields)
   {
     int idChild = (Integer) f.value;
     if (idChild != ID_NULL_NODE)
       ids.add(idChild);
   }
   return ids;
}


//This method returns null if it does not find a node.
private Object searchFixedNodeFromDB
 (Class<?> c, String fieldNameOfScalarField, Object value)
throws Exception
{
  String tableName =
    getFixedTableNameFromClassName(c.getName());

  if (!tableExists(tableName))
    return null;
  Integer instanceId = (Integer) readSingleValue
    (tableName,"instanceId", fieldNameOfScalarField+"=?",value);
  return readNodeFromDB(instanceId);
}
private Object readNodeFromDB(Integer id)
throws Exception
{
  if (id == ID_NULL_NODE)
    return null;

  //Test if the node has already been read.
  Object p = readNodes.get(id);
  if (p != null)
    return p;

  //Create the run-time node (object) corresponding the type of
  //the node in the object databse.
  p = createRunTimeNode(id);

  //Set that the node has been read now.
  readNodes.put(id,p);

  ArrayList<FieldT> fields = readFieldsOfDBNode(id,0);
  for(FieldT field : fields)
  {
    Object v = field.value;
    if (isPointerField(field.typeCode))
      v = readNodeFromDB((Integer) v); //Recursion.
    setValueInFieldOfObject(p,v,field);
  }
  return p;
}


private void setValueInFieldOfObject
 (Object o, Object value, FieldT field)
throws Exception
{
  Class<?> c = o.getClass();
  if (c != ListNode.class)
```

```
  {
    Field f = (Field) field.field;
    f.set(o,value);
  }
  else
    ((ListNode)o).list.add(value);
}


Object createRunTimeNode(int id)
throws Exception
{
  Class<?> c = getClassOfDBNode(id);
  Object p;
  if (c != ListNode.class)
  {
    Constructor cnstr = c.getDeclaredConstructor();
    cnstr.setAccessible(true);
    p = cnstr.newInstance();
    //Set the id value of the run-time fixed node.
    setId(p,id);
  }
  else
  {
    ListNode list = new ListNode();
    //Set the id value of the run-time list node.
    list.id = id;
    p = list;
  }
  return p;
}


//Fields of the database node are returned in a ArrayList.
//Both a field and its content is returned in the structure
//FieldT. A caller selects with the selector if only scalar
//fields or only pointer fields, or if both types of fields
//are returned.
//
//The id defines the node in the database.
//
//Values of the selector:
//0 = Pointer fields and scalar fields are returned.
//1 = Only pointer fields are returned.
//2 = Only scalar fields are returned.
ArrayList<FieldT> readFieldsOfDBNode(int id, int selector)
throws Exception
{
  Class<?> c = getClassOfDBNode(id);
  if (c != ListNode.class)
    return readFieldsOfDBFixedNode(c,id,selector);
  else
    return readFieldsOfDBListNode(id,selector);
}
private ArrayList<FieldT> readFieldsOfDBFixedNode
  (Class<?> c, int id, int selector)
throws Exception
{
  String tableName =
    getFixedTableNameFromClassName(c.getName());

  int rowId = (Integer) readSingleValue
    (tableName,"id","instanceId=?",id);

  //Get the fields from the corresponding run-time node type
  //(class). Do not fill the fields with any values yet.
  ArrayList<FieldT> fields = getFields(c,null,selector);

  //Fill the fields from the database.
  for(FieldT f : fields)
  {
```

```
      String fieldName = ((Field) f.field).getName();
      //Set the value in the field. The value is read from the
      //database.
      f.value = readSingleValue(tableName,fieldName,"id=?",rowId);
    }
    return fields;
  }
  private ArrayList<FieldT> readFieldsOfDBListNode
    (int id, int selector)
  throws Exception
  {
    ArrayList<FieldT> fields = new ArrayList<FieldT>();

    int rowIdOfList = (Integer) readSingleValue
      ("lists","id","instanceId=?",id);

    int len = (Integer) readSingleValue
      ("lists","len","id=?",rowIdOfList);

    for(int pos=0; pos<len; ++pos)
    {
      int typeCodeOfField = (Integer) readSingleValue
        ("listItems", "type", "parent=? AND position=?",
         rowIdOfList,pos);

      boolean addPointerField =
        isPointerField(typeCodeOfField) &&
          (selector == 0 || selector == 1);

      boolean addScalarField =
        isScalarField(typeCodeOfField) &&
          (selector == 0 || selector == 2);

      if (addPointerField || addScalarField)
      {
        Object value = readSingleValue
          ("listItems", "item", "parent=? AND position=?",
           rowIdOfList,pos);

        fields.add(new FieldT(pos,typeCodeOfField,value));
      }
    }
    return fields;
  }


//This method creates the table for a fixed node (for its type).
//These tables are created dynamically, when the system encounters
//the type of the fixed node for the first time.
//
//Below are two examples of Java types of fixed nodes, Friend and
//Book. Fields of the these nodes (their Java types) are public,
//but this is not a requirement. Every node (object) in the
//run-time memory contains a compulsory id field.
//
//
//  package userclasses;
//
//  public class Friend
//  {
//    //Id of the node. If id==0 the node does not have a
//    //corresponding node in the database. If id>0 the node has
//    //a corresponding node in the database identified by the
//    //id.
//    public int id;
//
//    //Scalar field of type String.
//    public String name;
//
//    //Scalar field of type Integer.
//    public Integer age;
```

```
//
//    //Pointer field referring to a list node. Contains Java
//    //reference to an object of type ListNode (or null).
//    public ListNode list;
//
//    //Pointer field referring to a fixed node of type Book.
//    //Contains Java reference to an object of type Book
//    (or null).
//    public Book book;
//
//    //Default constructor is required. The database uses
//    //this constructor to create a corresponding Java object
//    //when a node is loaded from the database with the search
//    //method.
//    public Friend(){}
//
//    <Possible other methods>
//  }
//
//  public class Book
//  {
//    public int id;
//
//    public String name;
//    public int price;
//
//    public Book(){}
//
//    <Possible other methods>
//  }
//
//
//Database tables created for the types of fixed nodes above:
//
//
//  CREATE TABLE userclasses_Friend
//  (
//    //Primary key.
//    id INTEGER PRIMARY KEY AUTOINCREMENT,
//
//    //Refers to nodeInstances.id.
//    instanceId INTEGER,
//
//    //Scalar field.
//    name TEXT,
//
//    //Scalar field.
//    age INTEGER,
//
//    //Pointer field referring to a list node. Refers to
//    //nodeInstances.id
//    //(or contains zero if refers to a null list node).
//    list INTEGER,
//
//    //Pointer field referring to a fixed node of type book.
//    //Refers to nodeInstances.id
//    //(or contains zero if refers to a null book node).
//    book INTEGER
//  )
//
//  CREATE TABLE userclasses_Book
//  (
//    id INTEGER PRIMARY KEY AUTOINCREMENT,
//    instanceId INTEGER,
//    name TEXT,
//    price INTEGER
//  )
  private void createDBTableForFixedNode(Class<?> c)
  throws Exception
  {
    String tableName =
```

```
    getFixedTableNameFromClassName(c.getName());

  String str = "CREATE TABLE "+tableName
             +" ("
             +"id INTEGER PRIMARY KEY AUTOINCREMENT"
             +",instanceId INTEGER";

  ArrayList<FieldT> fields = getFields(c,null,0);
  for (FieldT f : fields)
  {
    String fieldName = ((Field) f.field).getName();

    String sqlFieldType =
      getSQLFixedNodeFieldTypeFromTypeCode(f.typeCode);

    str += ("," + fieldName + " " + sqlFieldType);
  }
  str += ")";
  executeStatement(str);
}


private int readORC(int id)
throws Exception
{
   return (Integer)
     readSingleValue("nodeInstances","orc" , "id=?",id);
}
private int readIRC(int id)
throws Exception
{
   return (Integer)
     readSingleValue("nodeInstances","irc" , "id=?",id);
}
private void incrORC(int id)
throws Exception
{
  executeStatement
    ("UPDATE nodeInstances SET orc=orc+1 WHERE id=?",id);
}
private void decrORC(int id)
throws Exception
{
  executeStatement
    ("UPDATE nodeInstances SET orc=orc-1 WHERE id=?",id);
}
private void incrIRC(int id)
throws Exception
{
  executeStatement
     ("UPDATE nodeInstances SET irc=irc+1 WHERE id=?",id);
}
private void decrIRC(int id)
throws Exception
{
  executeStatement
    ("UPDATE nodeInstances SET irc=irc-1 WHERE id=?",id);
}


private static String
  getFixedTableNameFromClassName(String className)
{
  return className.replace('.','_');
}


private Class<?> getClassOfDBNode(int id)
throws Exception
{
  String className = (String)
```

```
      readSingleValue("nodeInstances","className","id=?",id);
    return  Class.forName(className);
  }


  private boolean tableExists(String tableName)
  throws Exception
  {
    int count  = (Integer)
      readSingleValue("sqlite_master",
                      "count(*)",
                      "type=? AND name=?",
                      "table",
                      tableName);

    return (count == 1);
  }
// Tool methods for the nodes in DB.
/////////////////////////////////////////////////////////////////


/////////////////////////////////////////////////////////////////
// Lower level tool methods.
  private static boolean isPointerField(int typeCodeOfField)
  {
    return typeCodeOfField == FIELD_TYPE_FIXED_NODE
    || typeCodeOfField == FIELD_TYPE_LIST_NODE;
  }
  private static boolean isScalarField(int typeCodeOfField)
  {
    return typeCodeOfField == FIELD_TYPE_INT
    || typeCodeOfField == FIELD_TYPE_INTEGER
    || typeCodeOfField == FIELD_TYPE_STRING
    || typeCodeOfField == FIELD_TYPE_NONE;
  }


  private static int getTypeCodeOfFieldFromValueInField
    (Object value)
  {
    if (value==null)
      return FIELD_TYPE_NONE;
   Class<?> cf = value.getClass();
   if (cf == Integer.class) return FIELD_TYPE_INTEGER;
   else if (cf == String.class) return FIELD_TYPE_STRING;
   else if (cf == ListNode.class) return FIELD_TYPE_LIST_NODE;
   else return FIELD_TYPE_FIXED_NODE;
  }


  private static int getTypeCodeOfFieldFromClass(Class<?> cf)
  {
    if (cf == int.class) return FIELD_TYPE_INT;
    else if (cf == Integer.class) return FIELD_TYPE_INTEGER;
    else if (cf == String.class) return FIELD_TYPE_STRING;
    else if (cf == ListNode.class) return FIELD_TYPE_LIST_NODE;
    else return FIELD_TYPE_FIXED_NODE;
  }

  private static String getSQLFixedNodeFieldTypeFromTypeCode
    (int typeCode)
  {
    if (typeCode == FIELD_TYPE_INT) return "INTEGER";
    else if (typeCode == FIELD_TYPE_INTEGER) return "INTEGER";
    else if (typeCode == FIELD_TYPE_STRING) return "TEXT";
    else if (typeCode == FIELD_TYPE_FIXED_NODE) return "INTEGER";
    else if (typeCode == FIELD_TYPE_LIST_NODE) return "INTEGER";
    return null; //Should not happen.
  }

// Lower level tool methods.
```

```
////////////////////////////////////////////////////////////////////


////////////////////////////////////////////////////////////////////
// Lower level SQL methods.
  //Returns null if nothing found.
  private Object readSingleValue(String table,
                                 String field,
                                 String wherePart,
                                 Object... parameters)
  throws Exception
  {
    PreparedStatement st = getPrepStatement
      ("SELECT "+field+" FROM "+table+" WHERE "+wherePart,
       parameters);

    ResultSet rs = st.executeQuery();

    Object obj = null;
    boolean found = rs.next();
    if (found)
      obj = rs.getObject(1);
    st.close();
    return obj;
  }


  private void updateSingleValue(String table,
                                 String field,
                                 String wherePart,
                                 Object... parameters)
  throws Exception
  {
    executeStatement
      ("UPDATE "+table+" SET "+field+"=? WHERE "+wherePart,
        parameters);
  }


  private int doInsertReturnPrimaryKey(String str,
                                       Object... parameters)
  throws Exception
  {
    PreparedStatement st = getPrepStatement(str, parameters);
    st.executeUpdate();
    ResultSet rs = st.getGeneratedKeys();
    rs.next();
    int id = rs.getInt(1);
    rs.close();
    st.close();
    return id;
  }

  private void executeDelete(String table,
                             String wherePart,
                             Object... parameters)
  throws Exception
  {
    executeStatement
      ("DELETE FROM "+table+" WHERE "+wherePart, parameters);
  }

  private void executeStatement(String str, Object... parameters)
  throws Exception
  {
    PreparedStatement st = getPrepStatement(str, parameters);
    st.executeUpdate();
    st.close();
  }

  private PreparedStatement getPrepStatement(String str,
```

```
                                          Object... parameters)
  throws Exception
  {
    PreparedStatement st = connection.prepareStatement(str);
    for(int i=0; i<parameters.length; ++i)
      st.setObject(i+1,parameters[i]);
    return st;
  }

  static private void p(Object o)
  {
    System.out.println(""+o);
  }
// Lower level SQL methods.
//////////////////////////////////////////////////////////////////
}
```

## 9.2 File ListNode.java

```
/*
 * Copyright (c) 2016 Heikki Virkkunen.
 * Date: 5 April 2016
*/

package fi.heolvi.embed.base;

import java.util.ArrayList;
import java.lang.reflect.*;

public class ListNode
{

  //Id of a list node. If id==0 the list node does not
  //have a corresponding node in the database.
  //If id>0 the list node has a corresponding node in the
  //database identified by the id.
  public int id;

  //ArrayList containing items of list.
  public ArrayList<Object> list = new ArrayList<Object>();

  //Default constructor. The database uses
  //this constructor to create a corresponding Java object
  //when a list node is loaded from the database with the
  //search method.
  public ListNode(){}

  //Helping wrapper methods for ArrayList.
  public void add(Object o)
  {
    list.add(o);
  }
  public Object get(int index)
  {
    return list.get(index);
  }
  public Object remove(int index)
  {
    return list.remove(index);
  }
  public Object set(int index, Object element)
  {
    return list.set(index,element);
  }

  public String toString()
  {
    if (list==null)
      return "null";
```

```
    String s = "";
    s+="[";
    int i = 0;
    for(Object o:list)
    {
      if (i>0)
        s += ",";
      ++i;

      if (o==null)
        s+="null";
      else if (o.getClass() == Integer.class  || o.getClass() == Float.class)
        s+=o;
      else if (o.getClass() == String.class)
        s+=o;
      else if (o.getClass() == ListNode.class)
        s += "size=" + ((ListNode)o).list.size();
      else
      {
       String name = null;
       try
       {
         Field f = o.getClass().getField("name");
         name = (String) f.get(o);
       }
       catch (Exception e)
       {
       }
       s += name;
      }


    }//for
    s += "]";

    return s;
  }
}
```

## 9.3 File FieldT.java

```
/*
 * Copyright (c) 2016 Heikki Virkkunen.
 * Date: 5 March 2016
*/

package fi.heolvi.embed.base;

import java.lang.reflect.*;

//This is an abstraction for a field of a node.
//Used for fixed nodes and list nodes.
//Used for run-time nodes and nodes in the object database.
class FieldT
{
  int typeCode;  //Typecode of a field. Defines type of data in
                 //a field.
  Object field;  //For a fixed node java.lang.reflect.Field.
                 //For a list node position (integer 0,1,..) in
                 //a list.

  Object value;  //Value in a field.

  FieldT(Field f, int typeCode, Object value)
  {
    this.typeCode = typeCode;
    this.field = f;
    this.value = value;
  }
```

```
  FieldT(Integer pos, int typeCode, Object value)
  {
    this.typeCode = typeCode;
    this.field = pos;
    this.value = value;
  }

  public String toString()
  {
    return
    "FieldT"

    +"["

    +"field="+field

    +","
    +"isFixedNode="+typeCode

    +","
    +"value="+value

    +"]"
    ;
  }

}
```